\documentclass[floatfix,prb,preprint,amsmath,amssymb,aps,longbibliography]{revtex4-2}
\usepackage{graphicx}
\usepackage{amsfonts}  
\usepackage{amsmath}   
\usepackage{amssymb}   
\usepackage{color}
\usepackage{bm}
\usepackage{array}
\usepackage{ascmac}
\usepackage{fancybox}
\usepackage[version=4]{mhchem}
\usepackage{hyperref}
\usepackage{booktabs}
\hypersetup{
    colorlinks=true,
    linkcolor=blue,
    filecolor=magenta,      
    urlcolor=blue,
    citecolor=blue,
}
\usepackage{tabularx}
\usepackage{mathtools}
\usepackage{booktabs}
\usepackage{multirow}
\usepackage[colorinlistoftodos]{todonotes}
\usepackage{soul}
\usepackage{enumitem}

\begin{document}



\title{An ab initio approach to energy alignment and charge-state prediction of adsorbates on ultrathin insulators}

\author{Kevin Lizárraga$^{1,2}$}
\author{Saba Taherpour$^{1,3}$}
\author{Cesar E. P. Villegas $^{4}$}
\author{Christoph Wolf$^{1,2}$}

\affiliation{$^{1}$Center for Quantum Nanoscience, Institute for Basic Science (IBS), Seoul 03760, Republic of Korea}
\affiliation{$^{2}$Ewha Womans University, Seoul 03760, Republic of Korea}
\affiliation{$^{3}$Department of Physics, Ewha Womans University, Seoul 03760, Republic of Korea}
\affiliation{$^{4}$Departamento de Ciencias, Universidad Privada del Norte, Lima 15434, Peru.}

\begin{abstract}  
The rapid progress of electron spin resonance scanning tunneling microscopy experiments has enabled the manipulation of individual adsorbate spin states physisorbed on ultrathin oxide layers supported on metal substrates. Electron resonance requires unpaired spin density on the adsorbate, which can be achieved, for instance, through charge transfer from the supporting substrate. This requires the correct energy-level alignment between the energy levels of the adsorbate and the Fermi energy of the substrate. Experiments on molecules and single atoms adsorbed on metal-insulator systems have revealed complex phenomena, including electronic bandgap narrowing, charge transfer, Fermi-level pinning, and the re-ordering of adsorbate orbitals after charge transfer. Despite these advances, a predictive first-principles approach based on accurate methods such as quasiparticle GW, capable of capturing these effects without the prohibitive cost of full adsorbate/oxide/metal simulations, remains an open challenge. In this work, we present a theoretical approach to determine the energy-level alignment of adsorbates on oxide/metal substrates. Our method transparently exposes all physical processes and strikes a balance between computational cost and accuracy. Ionization potentials and electron affinities of the isolated adsorbates are obtained using GW calculations, electronic bandgap polarization is quantified through the quasiparticle renormalization caused by the substrate, Fermi-level pinning is evaluated within the integer charge transfer model, and work function shifts arising from Pauli pushback or from the adsorbate-metal dipole are determined from the local variations of the electrostatic potential. This computationally efficient framework paves the way for high-throughput screening of molecular qubits and organic electronic interfaces.
\end{abstract}

\maketitle
\begin{figure}[!htb]
    \centering\includegraphics[width=1.0\linewidth]{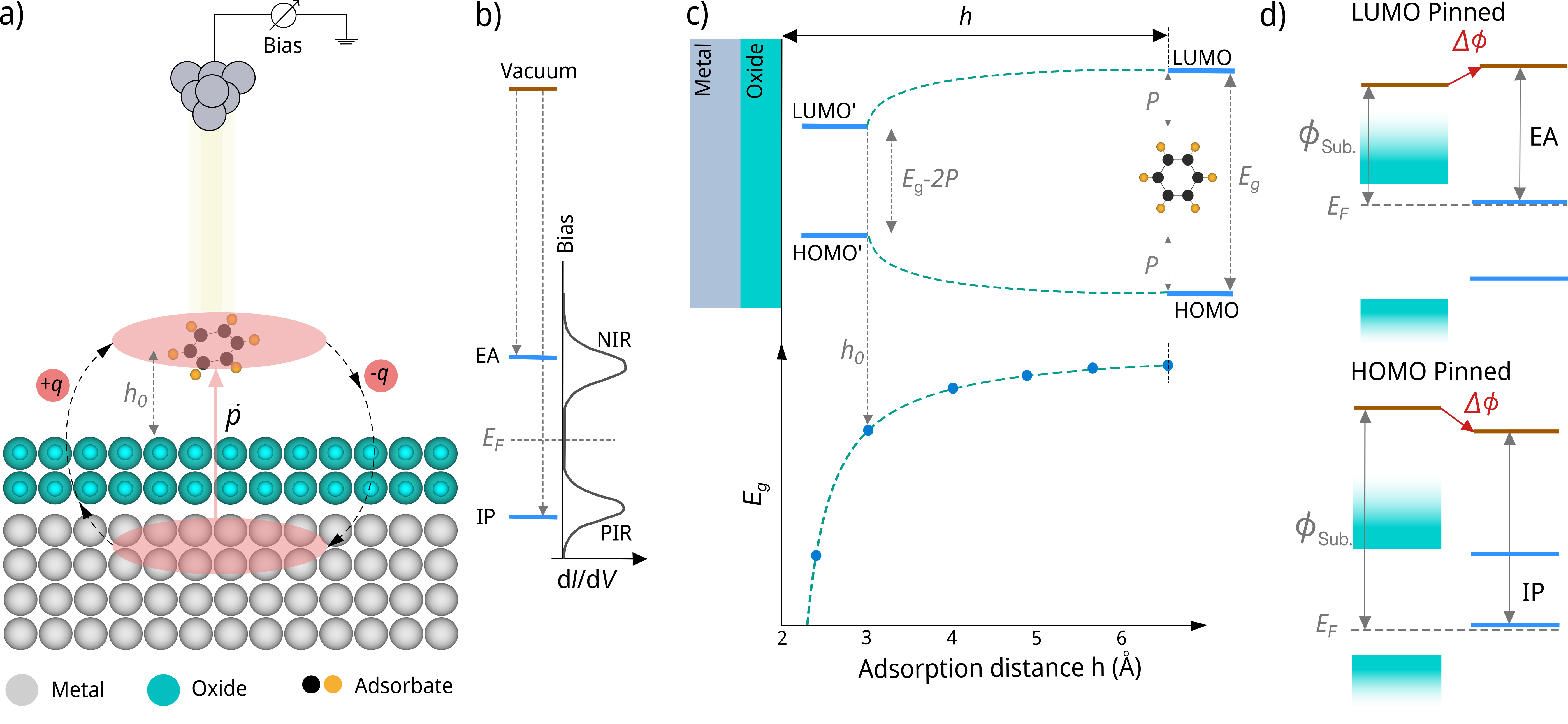}
    \caption{(a) Schematic of a STM measurement probing the electronic levels of a benzene molecule  adsorbed on an oxide/metal substrate. Upon adsorption, the molecule might undergo charge transfer, indicated by $\pm q$, which causes the formation of an electric dipole $\vec{p}$. (b) Position of the energy levels, electron affinity (EA) and ionization potential (IP), probed by an STS measurement of the d$I$/d$V$, yielding positive (negative) ion resonances (PIR, NIR) when the external bias matches the energies of removing (adding) an electron. (c) Distance-dependent polarization $P$ due to screening of the HOMO and LUMO energy levels. The polarized LUMO$'$ and HOMO$'$ are separated by an electronic bandgap $E_g$ that shrinks by $2P$ as the adsorption height $h$ decreases, approaching the adsorption distance $h_0$. (d) Charge-transfer regimes depending on the direction of electron flow. When an electron transfers from the substrate to the adsorbate, the LUMO becomes pinned to the Fermi level, leading to an increase in the substrate work function $\phi_\mathrm{sub}$ by $\Delta\phi$. Conversely, when charge transfer occurs from the adsorbate to the substrate, the HOMO becomes pinned, resulting in a decrease in the work function by $\Delta\phi$.}
    \label{fig:schematic}
\end{figure}

 
Spin-bearing adsorbates such as atoms or molecules on surfaces represent a versatile class of building blocks for the bottom-up design of atomic-scale quantum-coherent systems~\cite{Heinrich2021,Chen_2022, Maiellaro2024, Wolf_2024}. By tailoring the interactions between substrate and adsorbates, their charge and spin states can be engineered for a desired outcome such as $S=1/2$ quantum bit (qubit), $S>1/2$ quantum d-ary digits (qudit) or $S=1$ Haldane chains~\cite{Baumann_2015,mullegger_2015,Yang_2019,Natterer_2017, Choi_2025, Guo_2026}. To determine the final charge and spin state of an adsorbate, it is essential to understand the physical processes governing its interaction with the substrate. In this context, ultrathin oxide layers supported on metal substrates have become the prototypical platforms for quantum-coherent systems~\cite{Liu_2014}. These architectures emerge because the thin oxide layer acts as an electronic decoupling barrier that preserves the adsorbate orbital structure, promotes physisorption over chemisorption~\cite{Kawaguchi2023,Choi_2025}, and enables the emergence of well-defined Kondo signatures when unpaired electrons are present in the adsorbate~\cite{Guo_2026,Wolf_2024}. At the same time, the metal enables charge transfer through tunneling, allowing neutral molecules like pentacene to acquire unpaired spin~\cite{Hollerer2017}. Among these systems, adsorbates like molecules and single atoms have been deposited on top of few-layer binary oxides such as magnesium oxide (MgO) or sodium chloride (NaCl) films grown on metallic substrates (such as Ag, Au, or Cu)~\cite{Guo_2026}. A schematic of this architecture along with a representative alignment of the frontier orbitals is shown in Fig.~\ref{fig:schematic}a and b. Representative molecular adsorbates include perylenetetracarboxylic dianhydride (PTCDA)~\cite{Schaal_2024,Hurdax_2025}, terbium(III) bisphthalocyanine (TbPc$_2$)~\cite{Kawaguchi2023}, iron(II)phthalocyanine (FePc)~\cite{Zhang_2023,Willke_2021,Colazzo_2025}, pentacene (pentacene)~\cite{Hollerer2017,kovarik2024}, polycyclic aromatic molecules~\cite{czap2025magneticresonanceimagingsingle}, and Tetracyanoethylene (TCNE)~\cite{Zhang_2020}. Single-atom systems include transition metals such as Ti~\cite{Yang_2017} and Fe~\cite{Willke_2018,Seifert_2020}, as well as rare-earth elements including Eu~\cite{Czap_2025}, Er~\cite{Reale_2024}, and Tm~\cite{Reale2023PRB}.

Previous studies of adsorbates on oxide/metal interfaces have identified several mechanisms, such as the Pauli pushback effect, electronic polarization, Fermi level pinning, and the rearrangement of molecular orbitals after charge transfer~\cite{Greiner_2011,Greiner_2013,Ley_2012,Li_2020,Chai_2014,Hollerer2017,Schaal_2024}. The electronic polarization of the band gap resulting from changes in the electron affinity (EA) and ionization potential (IP)—after the impurity is adsorbed at a height $h_0$ (see Fig.~\ref{fig:schematic}c)—has been reported in Refs.~\cite{neaton2006,lastra2009}. However, these works typically consider only a single substrate such as either an oxide or a metal thereby neglecting the effects of the oxide/metal interfaces. In addition, $non-local$ techniques like ultraviolet photoelectron spectroscopy (UPS) studies have revealed the pinning of either the highest occupied molecular orbital (HOMO) or lowest unoccupied molecular orbital (LUMO), indicating a universal integer charge transfer behavior with negative or positive work function shifts, respectively, of $\sim0.3$~eV (see Fig.~\ref{fig:schematic}d)~\cite{Chai_2014,Ley_2012}. In contrast, this shift increases to $\sim1$~eV when $local$ probes, such as scanning tunneling microscopy and spectroscopy (STM/STS) are used, as reported for PTCDA and pentacene~\cite{Hollerer2017,Schaal_2024}. Nonetheless, these models typically predict work function shifts on the order of $\sim0.3$ eV, which is in stark contrast to the experimentally observed shifts of around $\sim 1$ eV for PTCDA, pentacene, and FePc deposited on oxide/metal interfaces probed with highly local spectroscopic techniques such as scanning tunneling microscopy and spectroscopy (STM, STS) \cite{Hollerer2017,Schaal_2024}. Moreover, re-ordering of the molecular orbitals after the ionization has been suggested in Refs.~\cite{Hollerer2017,Schaal_2024}. These studies combined experiments and simulations to reconstruct the final molecular orbital alignment after adsorption. A study based solely on ab initio methods is still lacking. 

This work presents a reliable theoretical method, based on density functional theory (DFT) and many-body perturbation theory within the GW approach, for predicting the energy alignment of impurities on oxide/metal substrates. The main advantages of our approach are: (i) the separation of the key physical mechanisms governing adsorption, enabling clearer interpretation; (ii) its general applicability across a wide range of adsorbates and substrates, demonstrating robust predictive capability; and (iii) the avoidance of full interface simulations with the high-demanding GW method, thereby reducing computational cost while retaining quantitative reliability. We use DFT calculations within the generalized gradient approximation (GGA) to describe the Pauli pushback effect, the work function of the substrates and the electric-dipole shifts of the charge transfer. The quasiparticle GW method is used to obtain reliable chemical potentials, polarization effects, and the correct reordering of molecular orbitals, which are typically underestimated at lower levels of theory. Finally, we also predict the charge injection barriers between adsorbates and electrodes that control quantum coherence during transport.

\section{Results}
\subsection{Determination of Ionization Potential and Electron Affinities }

The estimation of the energy level alignment begins by determining the positions of the HOMO and LUMO energy levels with respect to the vacuum level. We use Koopman's theorem in the GW approximation which relates the position of the HOMO to the cost of removing one electron ($N\rightarrow N-1$), i.e. the ionization potential (IP), and the LUMO to the cost of adding an electron to the adsorbate ($N\rightarrow N+1$) which is the electron affinity (EA)~\cite{Forster2024GW}. The difference is the fundamental electronic gap $E_g= \mathrm{IP}-\mathrm{EA}$~\cite{kang2016}. In the GW method IP and EA with respect to the vacuum level $V$ are given by~\cite{huser2013,blase2011,kang2016,rostgaard2010,bruneval2012,caruso2012}:

\begin{align}
    \mathrm{IP}&=-(E^{V}_\mathrm{HOMO}) \nonumber \\
    \mathrm{EA}&=-(E^{V}_\mathrm{LUMO}).
\label{eq:GW_IP}
\end{align}

We used the COHSEX approximation of the GW \cite{neaton2006,huser2013,masoud_2021,onida2002} to calculate the corrected eigenenergies $E^{V}_\mathrm{HOMO}, E^{V}_\mathrm{LUMO}$ as it has been shown that COHSEX gives improved values for EA and IP when compared to single-shot G$_0$W$_0$ and the self-consistent GW$_0$ method for molecules~\cite{Knight_2016,Korbel_2014}. In our work, we separately simulate the isolated neutral and charged molecule to determine the $E^{V}_\mathrm{HOMO/LUMO}$ energies in the case of neutral molecules and $E^{V}_\mathrm{SOMO/SUMO}$ energies for singly charged molecules, which yields the unscreened electronic bandgap $E_g$ (the HOMO-LUMO gap) as shown in Fig.~\ref{fig:ips_and_eas}a.

\subsection{Work Function of Substrates}

\begin{figure}[!htb]
    \centering\includegraphics[width=1.0\linewidth]{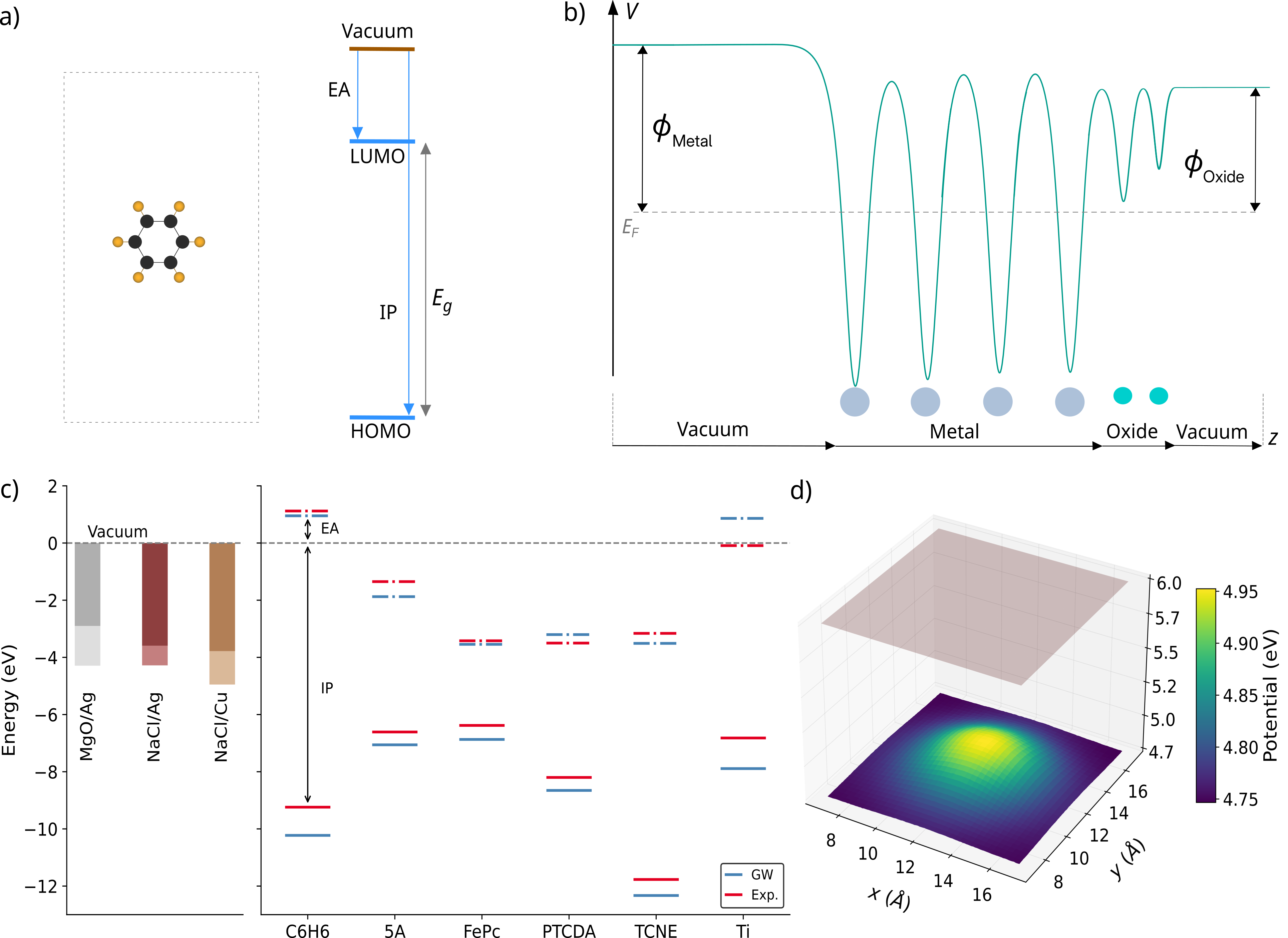}
    \caption{\textbf{Physical quantities governing the energy level alignment.} (a) Gas phase of a neutral molecule $[M]^0$ along with the LUMO and HOMO orbitals, that define the EA and IP with respect to the vacuum level, whose difference is the electronic bandgap $E_g$. (b) Electrostatic potential of oxide/metal system. The work function ($\phi$) is the difference between the vacuum and the Fermi energy. (c) left panel: Work functions of common substrates including the bare metal and oxide-modified values. Right panel: GW (blue) and experimental (red) values of the chemical potentials of EA and IP for the adsorbates. Dashed (solid) lines correspond to EA (IP). (d) Local electrostatic potential variation ($\Delta V$) for Benzene on MgO/Ag at a distance of 4~\AA ~above the molecular plane as probed by local scanning probe techniques.}
    \label{fig:ips_and_eas}
\end{figure}

The work function $\phi$ of a metal is the minimum energy required to remove an electron from the Fermi level of a material to the vacuum level. It reflects the surface dipole formed by the redistribution of electronic charge at the interface between the solid and the vacuum~\cite{Lang_1970,Luth_2001}. $\phi$ can be readily calculated from a slab-vacuum DFT calculation by taking the difference between the Fermi energy and the vacuum potential in a region far away from the surface (Fig.~\ref{fig:ips_and_eas}b). When the oxide layer is placed on the metal, the overlap of electronic densities causes Pauli repulsion ("push-back" effect) that pushes metal electron density back into the substrate, modifying the surface dipole and lowering the work function by $\Delta\phi_\textrm{PB}=\phi_\textrm{metal}-\phi_\textrm{oxide}$~\cite{Martinez_2008,Braun_2009,Hollerer2017}. In general, ultrathin oxide layers reduce the work function of metals, e.g., one, two and three monolayers (ML) of MgO reduce the work function of Ag(001) by up to $\Delta\phi_\textrm{PB}=1.3$~eV~\cite{Prada2008, Bieletzki2010, Baumann2014, Wolf2020}. Similarly, NaCl reduces the work function of Ag(111)~\cite{Wang_2021} or Cu(001)~\cite{Robledo2015}. All our oxide layers were calculated at ideal stoichiometry, however deviations from this stoichiometry in the experiment might cause further modifications of the work function~\cite{Nilius_2009, CHO2013541}. 

It is important to note that, the metal surfaces and their periodic vacuum regions must be separated by a sufficiently large distance to avoid artificial interactions between the periodic image in plane-wave DFT calculations which use periodic boundary conditions. Spurious interaction in finite vacuum cells can be corrected by adding a compensating sawtooth-like potential and a dipole in the vacuum region~\cite{Giannozzi_2009}. Since dispersive forces can influence the work function~\cite{Chiter2016} and may be relevant for the systems considered here, they are included self-consistently using the Grimme–D3 correction~\cite{Grimme_2010}.

\subsection{Bandgap Polarization}
\label{sec:gap_renorm}
The determination of the electronic bandgap using a molecule-in-a-box system corresponds to the molecule in perfect vacuum ($\varepsilon_r=1$). However, $E_g$ is sensitive to the dielectric environment of the substrate~\cite{lastra2009,Hollerer2017,Schaal_2024}. When the molecule is adsorbed on an oxide or metal substrate the effective dielectric constant $\varepsilon_r>1$ lowers the LUMO energy and raises the HOMO energy by the polarization energy $P$, thereby reducing the fundamental electronic bandgap $E_g\rightarrow E_g'=E_g-2P$~\cite{neaton2006,lastra2009}. $P$ strongly depends on the adsorption height $h$, as shown in Fig.~\ref{fig:schematic}c. Previous studies show that $P$ depend on the static exchange contribution to the self-energy, which are well accounted for the GW method~\cite{neaton2006,lastra2009}. In contrast, using generalized gradient approximation (GGA) or hybrid functionals such as PBE0 show an underestimation of the bandgap and an inability to differentiate the chemical composition of the substrate e.g. producing similar values for metals and oxides~\cite{lastra2009}. To account for the screening effect, we calculate the quasiparticle energies of the molecule adsorbed on the oxide layer at the equilibrium adsorption height $h_0$ from a relaxed DFT simulation. Lastly, the geometry of neutral and charged molecules can differ slightly. The inner reorganization energy $\lambda=E^{-}(Q_0)-E^{-}(Q)$, i.e. the energy difference between the charged molecule at the geometry of the neutral molecule $(Q_0)$ and the fully relaxed charged molecule $(Q)$ is calculated at the GGA level~\cite{Coropceanu2007}.

\subsection{Charge Transfer and Fermi Level Pinning}
\label{sec:CT_EF_pinning}
Charge transfer in an adsorbate/substrate system can occur either from the substrate to the adsorbate or vice versa when the renormalized LUMO or HOMO at $h_0$ is close to the Fermi energy~\cite{Greiner_2011,Chai_2014,Ley_2012}. In the absence of charge transfer, the combined adsorbate-substrate system ($\phi_{\textrm{sub,ads}}$) follows the classical Schottky–Mott rule~\cite{Luth_2001,Ley_2012}, such that $\phi_{\textrm{sub,ads}}=\phi_\mathrm{\textrm{sub}}$. Consequently, variations in the substrate work function ($\Delta\phi_\mathrm{\textrm{sub}}$) for a fixed organic overlayer produce an equivalent shift in the adsorbate/substrate system $\Delta\phi_{\textrm{sub,ads}}$~\cite{Luth_2001,Ley_2012}.

\subsubsection{Fermi Level Pinning}
On the other hand, when a renormalized LUMO or HOMO approaches the Fermi level, charge transfer occurs, inducing an image charge in the metallic substrate~\cite{Greiner_2011,Chai_2014,Ley_2012}. The changes in the electric potential produced by this dipole lead to a shift of the vacuum level, $\Delta\phi$~\cite{Frondelius2008, Giordano2011, Hollerer2017}. When the LUMO (HOMO) level becomes pinned, $\Delta\phi$ increases (decreases), leading to a corresponding upward (downward) shift of the vacuum level, as shown in Fig.~\ref{fig:schematic}d. In the highly diluted limit relevant for scanning probes, this shift can be calculated from the \textit{local} variation of the potential of the adsorbate/oxide/metal system (Fig.~\ref{fig:ips_and_eas}d, see also Figs.~\ref{fig:charge_transfer}, \ref{fig:WF_height}). For surface averaging techniques and dense molecular layers $\Delta\phi$ is well described by a Boltzmann-averaged capacitor model proposed in Refs.~\cite{Greiner_2011,Ley_2012,Yang_2017,Oehzelt_2014,Chai_2014}.

\subsubsection{Local Potential Variations and Impact on Energy Level Alignment}
The measured work function shift, $\Delta\phi$, depends strongly on the experimental probe. In $\textit{local}$ measurements such as STM/STS, $\Delta\phi$ increases rapidly as the tip approaches the adsorbate, reflecting the strong sensitivity to the local electrostatic environment \cite{Yamada_2001}. In contrast, $\textit{non-local}$ techniques such as UPS and IPES probe much larger surface areas, yielding a spatially averaged $\Delta\phi$ over substrate, adsorbates, and defects \cite{Oehzelt_2014,Chai_2014}. This averaging typically reduces the apparent work function shift. Consequently, STM measurements often reveal larger work function shifts ($\Delta\phi \approx 0.8$~eV \cite{Hollerer2017,Schaal_2024}) than those obtained from UPS/IPES ($\Delta\phi \approx 0.2$~eV \cite{Greiner_2011}), highlighting the importance of spatial resolution.

\subsubsection{SUMO/SOMO Orbital splitting}
After charge transfer the molecular orbitals are re-organized. The energy level that gets pinned to the Fermi energy, splits into the Singly Occupied (Unoccupied) Molecular Orbital SOMO (SUMO)~\cite{Hollerer2017,Hofmann_2015,Hurdax_2025}. The energy difference between these orbitals is the screened electronic bandgap probed by local measurements such as scanning tunneling spectroscopy (STS) or photoemission experiments such as Ultraviolet Photoelectron Spectroscopy (UPS). Both energy levels can be directly obtained from GW simulations of the isolated charged molecule and the position of the energy levels can be directly compared to the d$I$/d$V$ maps obtained by STS measurements. We note that STS measurements are highly local probes whilst UPS measurements are surface-averaging, which can lead to position and coverage-dependent shifts of the vacuum level resulting in different alignment with the Fermi level of the substrate (see Fig.~\ref{fig:ips_and_eas}d). Importantly, for both $local$ and $non-local$ measurements, the electronic bandgap is affected by molecular aggregation~\cite{Cochrane_2015}.

Finally, we remark that if the ground-state of the adsorbate is effectively an electron spin $S=1/2$, the SOMO and SUMO of the adsorbate can be mapped to a single-orbital Anderson impurity model (SAIM)~\cite{Zhang_2025,reina2025, Urdaniz2025}. The SAIM has two relevant energy scales. The ionization energy $\varepsilon=\mathrm{IP}-E_\mathrm{F}$ to remove an electron relative to the Fermi level of the electrodes, and the Hubbard repulsion $U=\mathrm{IP}-\mathrm{EA}$ to add an additional electron.

\section{Discussion}

\subsection{Chemical Potentials of the Isolated Impurities}
GW values for IP and EA for isolated benzene (C$_6$H$_6$), pentacene, FePc, PTCDA, TCNE and Ti are presented in Table~\ref{tab:eaip} and Fig.~\ref{fig:ips_and_eas}c. These values are in good agreement with the experimental results~\cite{blase2011,huser2013,rostgaard2010,neaton2006,bruneval2012}.  Fig.~\ref{fig:ips_and_eas}c shows that the electron affinities of isolated FePc, PTCDA and TCNE molecules lie close to the work functions of common insulator/metal combinations (MgO/Ag, NaCl/Ag, and NaCl/Cu), indicating a tendency for electron transfer even at large molecule–substrate separations. In contrast, pentacene requires significant polarization $P>1$~eV to shift the LUMO toward the pinning regime.

\begin{table*}[!htb]
    \centering
    \begin{tabular*}{\textwidth}{@{\extracolsep{\fill}}c|cccccc}
    \hline
        Adsorbate & \multicolumn{2}{c}{GW}   & \multicolumn{2}{c}{$\Delta$SCF}   & \multicolumn{2}{c}{Experiment} \\
        \cline{2-3}\cline{4-5}\cline{6-7}
          & IP  &  EA  & IP & EA & IP & EA\\ \hline 
        C$_6$H$_6$ & 10.2 & -0.95 & 9.3 & -1.15& 9.24~\cite{Nemeth_1993} & -1.12~\cite{Burrow_1987} \\
        pentacene & 7.06 & 1.88 & 6.16 & 1.59& 6.61~\cite{Schmidt_1977} & 1.35~\cite{Crocker_1993} \\
        FePc &6.87 & 3.54 & 6.19 & 2.24 & 6.38~\cite{Berkowitz1979} & 3.42~\cite{isobe2023} \\
        PTCDA &8.65 & 3.20 & 7.93 & 3.07 & 8.20~\cite{dori2006} & 3.50~\cite{Hurdax_2025} \\
        TCNE & 12.33 & 3.51 & 11.52 & 3.21 & 11.77~\cite{Ikemoto_1974} & 3.16~\cite{Khuseynov_2012} \\
        \hline \hline
        Ti &7.89 & -0.86 & 7.73 & 0.08& 8.74~\cite{lide2003} & 0.09~\cite{NIST_ti} \\ \hline
    \end{tabular*}
    \caption{IP and EA calculated of molecules and a single Ti atom with COHSEX GW and the $\Delta$SCF method compared to experimental results. All values are in eV.}
    \label{tab:eaip}
\end{table*}

We also compare these values with values obtained from charge-constrained SCF calculations ($\Delta$SCF), in which we take the total energy difference of the $N$-electron system to the $N+1$ (anion) and $N-1$ (cation) systems~\cite{Park_2015,Kowalczyk_2011}. Generally, the chemical potentials agree well with the experiments (see Table~\ref{tab:eaip} and Fig.~\ref{fig:supp_deltascf_GW}). 

Importantly, Eq.~\eqref{eq:GW_IP} can be used to describe the frontier orbital energies of isolated charged systems ($M^{\pm q}$), allowing us to identify the SUMO and SOMO molecular orbitals after the charge transfer.

\subsection{Impurities Adsorbed on Oxide/Metal Substrates}

We benchmark our method to six representative systems studied by photoemission and scanning probe techniques. These are Benzene/NaCl/Cu~\cite{Robledo2015}, pentacene/MgO/Ag~\cite{Hollerer2017, kovarik2024}, FePc/MgO/Ag~\cite{Zhang_2023}, PTCDA/NaCl/Ag~\cite{Cochrane_2015}, TCNE/MgO/Ag~\cite{Zhang_2020} and Ti/MgO/Ag~\cite{Phark2026}. Our model is designed to satisfy the following criteria: (i) Determine an accurate electronic bandgap for the isolated adsorbate. (ii) Correctly estimate the polarization of the electronic bandgap. (iii) Predict the occurrence or lack of charge transfer along with the vacuum level shifts. (iv) Estimate the charge injection and extraction barriers.

\subsubsection{Benzene on NaCl/Cu}
\begin{figure}[!htb]
    \centering
    \includegraphics[width=0.60\linewidth]{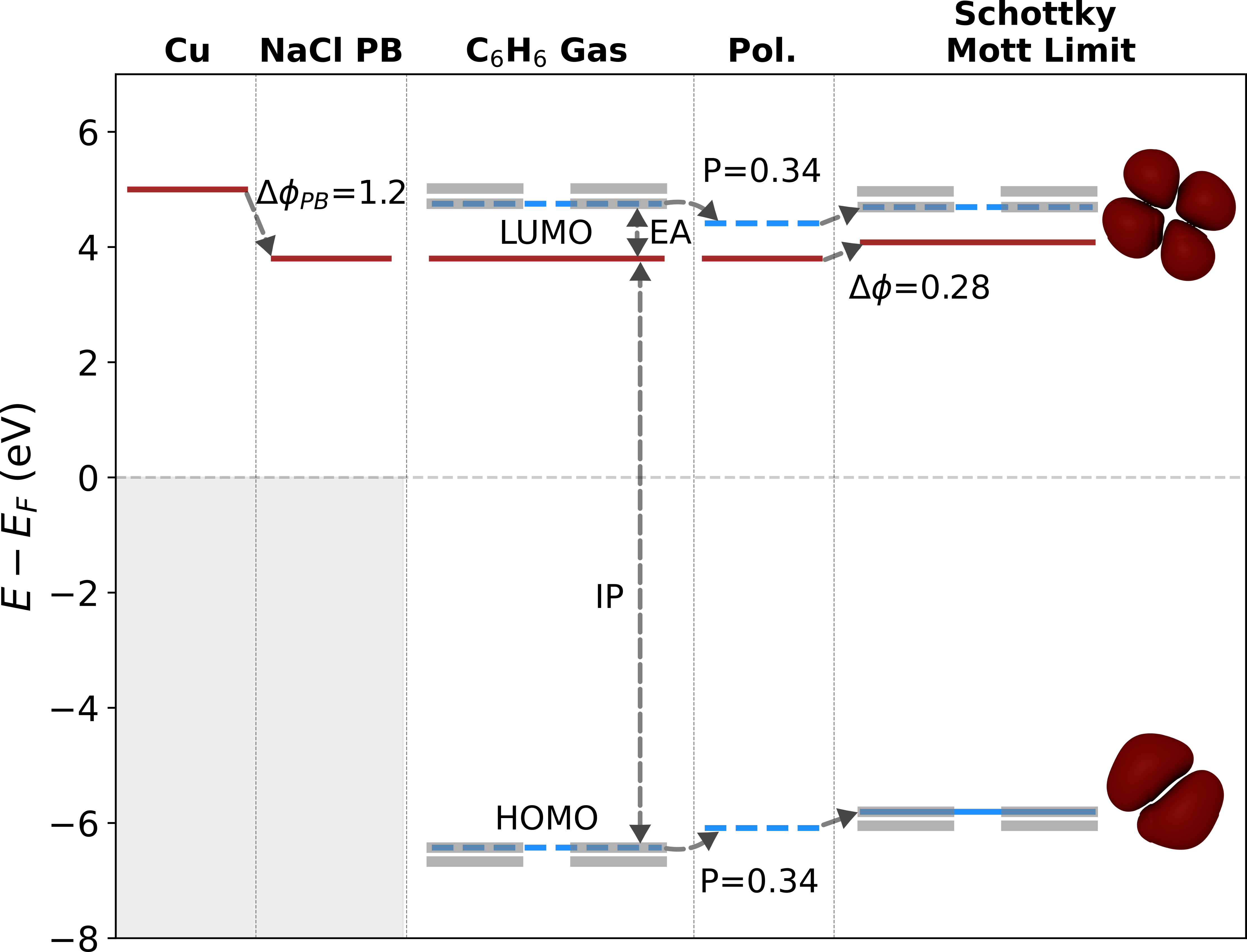}
    \caption{\textbf{Energy level alignment of Benzene on NaCl/Cu.} Each panel from left to right shows a different effect of the energy level alignment: Pauli pushback ($\Delta\phi_\textrm{PB}$), polarization $P$ (Pol.), and the  work function shift $\Delta\phi$, which here is in the Schottky-Mott limit. Grey lines show the molecular orbitals and blue dashed lines highlight the evolution of HOMO and LUMO. Molecular orbital densities of the frontier orbitals appear as red isosurfaces (iso$=6.7\times10^{-3}$e/A$^3$)}
    \label{fig:benzene}
\end{figure}

We start with the small closed shell organic molecule, benzene (C$_6$H$_6$). To determine the energy alignment of benzene on NaCl/Cu, we begin with the determination of the work function for NaCl/Cu(001). The work function of Cu(001) sets the vacuum level to 5.0~eV, whilst 2 ML of NaCl decreases the vacuum level by $\Delta\phi_\textrm{PB}=1.2$~eV, in good agreement with measurements of thin NaCl overlayers on Ag~\cite{Ploigt2007}. Then, we proceed to align the HOMO and LUMO levels with respect to the vacuum level of the system using the IP$=10.2$~eV and EA$=-0.95$~eV obtained from the GW calculation. The negative sign indicates that the anionic radical would be unstable without a polarizable medium. The substrate causes a screening of the electronic bandgap of $2P\approx 0.37$~eV on NaCl and 0.68~eV on NaCl/Cu. This reduces IP by $0.34$~eV and increases EA by $0.34$~eV, as can be seen in Fig.~\ref{fig:benzene}. After benzene is adsorbed on NaCl/Cu, the work function follows the Schottky-Mott limit by 0.28~eV~\cite{Luth_2001}. Finally, the renormalized EA and IP do not fall close to the Fermi Energy, suggesting the absence of a charge transfer mechanism. Our prediction of a neutral character of benzene on NaCl/Cu is in agreement with Ref.~\cite{Robledo2015}. To understand the role of the metal support we calculated the polarization of benzene on MgO/Ag ($2P=2.44$~eV) and NaCl/Ag ($2P=0.87$~eV). In all cases we found that adding the Ag layer adds $2\tilde{P}\approx0.5$~eV to the polarization (Fig.~\ref{fig:pol_benzene}), which will later be used to estimate the renormalization of system with large molecules for which our computational resources do not permit GW calculations of the combined metal/oxide/adsorbate system.

\begin{table*}[!htb]
    \centering
    \small 
    \setlength{\tabcolsep}{4pt} 
    \begin{tabular*}{\textwidth}{@{\extracolsep{\fill}}l|ccccc} 
        \toprule
        Adsorbate & $E_g$~(eV) & Metal & Oxide & $2P$~(eV) & $E'_g$~(eV) \\
        \midrule \midrule
        \multirow{5}{*}{C$_6$H$_6$} & \multirow{5}{*}{11.15} & -- & MgO & 1.95 & 9.20 \\
                                   &                        & Ag & MgO & 2.44 & 8.71 \\
                                   &                        & -- & NaCl & 0.37 & 10.78 \\
                                   &                        & Cu & NaCl & 0.68 & 10.47 \\
                                   &                        & Ag & NaCl & 0.87   & 10.28    \\
        \midrule
        \multirow{2}{*}{FePc}      & \multirow{2}{*}{3.33}  & -- & MgO & 0.67 & 2.66 \\
                                   &                        & Ag & MgO & 1.17$^*$ & 2.16 \\
        \midrule
        \multirow{2}{*}{pentacene}        & \multirow{2}{*}{5.18}  & -- & MgO & 0.59 & 4.59 \\
                                   &                        & Ag & MgO & 1.10$^*$ & 4.08 \\
        \midrule
        \multirow{2}{*}{PTCDA}     & \multirow{2}{*}{7.93}  & -- & NaCl & 0.28 & 7.66 \\
                                   &                        & Ag & NaCl & 0.78$^*$ & 7.16 \\
        \midrule
        TCNE                       & 11.52                  & Ag & MgO & 3.30 & 8.22 \\
        \midrule \midrule
        Ti                         & 8.74                   & Ag & MgO & 6.73 & 2.01 \\ 
        \bottomrule
    \end{tabular*}
    \caption{\textbf{Impact of screening on the electronic bandgap of the isolated adsorbates.} Unscreened electronic bandgap $E_g$ and polarization $2P$ of the adsorbates on oxide/metal substrates. $E'_g=E_g-2P$ is the screened electronic bandgap of the neutral molecule. $^* 2P$ here is the polarization obtained after adding the Ag contribution of $2\tilde{P}=0.5$~eV to the adsorbate/oxide polarization.}
    \label{tab:eaip2}
\end{table*}

\subsubsection{FePc on MgO/Ag}
\begin{figure}[!htb]
    \centering
        \includegraphics[width=1.00\linewidth]{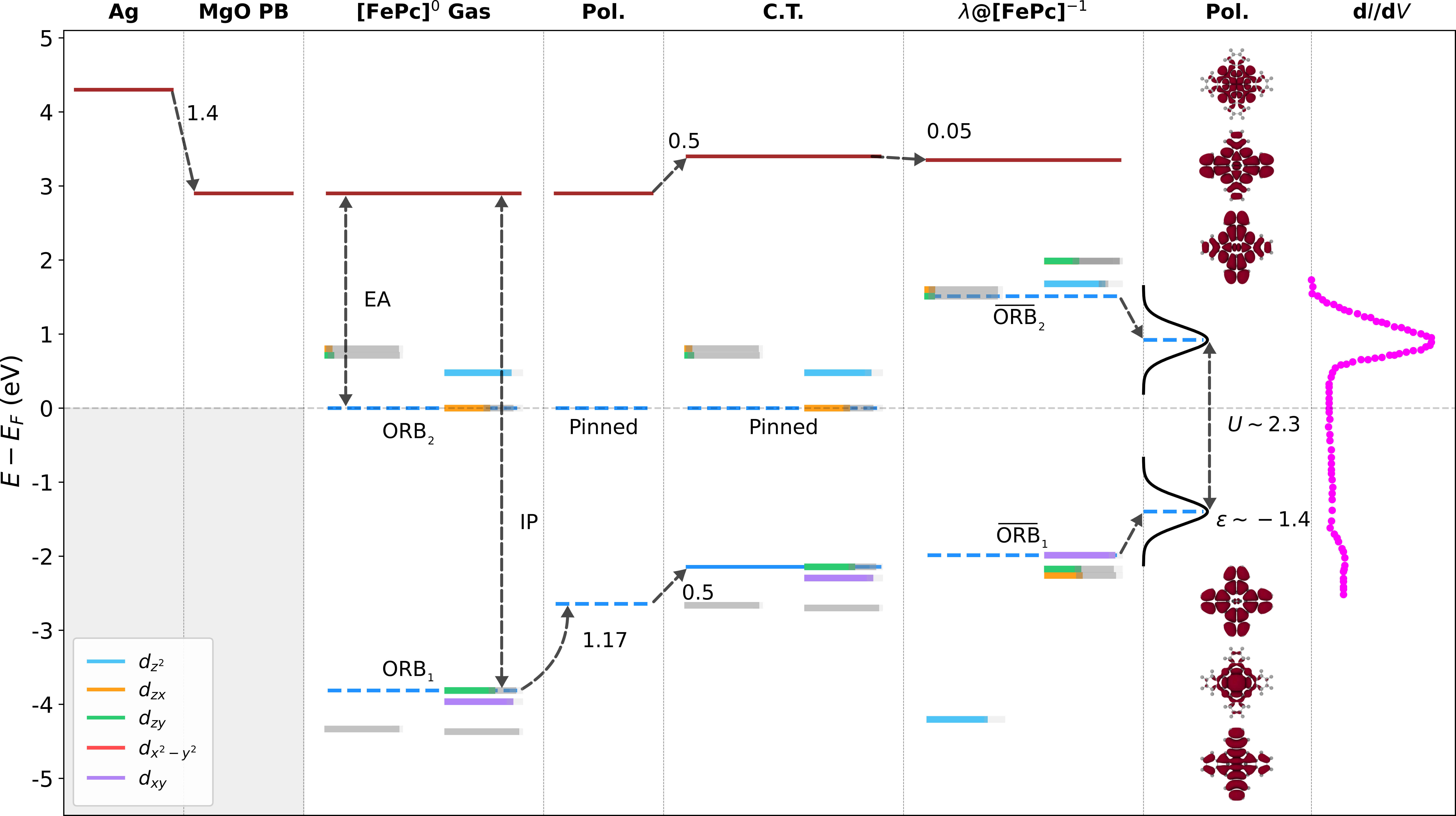} 
    \caption{\textbf{Energy level alignment of FePc on MgO/Ag.} The schematic from left to right correspond to all the effects such as Pauli pushback ($\Delta\phi_\textrm{PB}$), polarization $P$ (Pol.), charge transfer (C.T.), and reorganization of energy levels. The reordering of the molecular orbitals after charge transfer has been carried by considering the interaction of [FePc]$^{-1}$ with an axial water molecule that mimics the crystal field of the substrate. Frontier orbitals are shown as blue dashed lines and denoted as ORB$_1$ and ORB$_2$ for neutral FePc, and as $\overline{\mathrm{ORB}}_1$ and $\overline{\mathrm{ORB}}_2$ for the anion. Spin projection of the closest states above and below the frontier orbitals are plotted as continuum small colored bars. After the polarization of the [FePc]$^{-1}$ orbitals, the charging energy and barrier for charge extraction are denoted as $U$ and $\varepsilon$, respectively. Molecular orbital densities of the three first states below $\overline{\textrm{ORB}}_1$ and above $\overline{\textrm{ORB}}_2$ appear as red isosurfaces (iso$=6.7\times10^{-3}$e/A$^3$). Experimental data adapted from Ref.~\cite{Zhang_2025}. }
    \label{fig:fepc}
\end{figure}

Neutral Iron(II)phthalocyanine (FePc) is a 3$d^6$, $S=1$ system~\cite{rodriguez_2015,Urdaniz2025}. We align the vacuum level using the work function of the Ag(001) interface, $\phi=4.3$~eV, and account for two monolayers of MgO, which introduce a Pauli pushback of $\Delta\phi_\textrm{PB}=1.4$~eV. The 3$d$ transition metal core of FePc leads to a complicated orbital structure which could undergo significant reordering upon charging.\cite{Greulich2021} In the following, we focus on the alignment of the frontier orbitals which are the most relevant for charge transfer and the location of PIR and NIR. The position of ORB$_1$ and ORB$_2$ of the isolated molecule (IP$=6.87$~eV and EA$=3.54$~eV) relative to the Ag/MgO Fermi level already indicates the occurrence of charge transfer upon adsorption of the molecule. Consequently, substrate-induced screening reduces the electronic bandgap by approximately $2P\approx1.2$~eV. This value is the sum of polarization of FePc on MgO ($2P\approx 0.7$~eV) and an estimate of the contribution of silver ($2\tilde{P}\approx0.5$~eV). The silver contribution for benzene on MgO/Ag was calculated (see SI section \ref{sec:SI-benzene-pol}) because the number of atoms in the FePc/MgO/Ag system exceeds the computational resources of our GW calculation. Note that this correction could depend on the specific combination of metal/oxide and adsorbate. However, it yields reliable results for the molecules considered here, most likely due to their relatively simple flat structure, which resembles benzene. Due to pinning of ORB$_2$ at the Fermi level, the electronic bandgap shift is only observed in the ORB$_1$ level (see Fig.~\ref{fig:fepc}, columns 4 and 5). This explains why upon adsorption the substrate transfers a charge, i.e., [FePc]$^0$ $\rightarrow$ [FePc]$^{-1}$, in agreement with experiments~\cite{Zhang_2023,Willke_2021,Zhang_2025,greule2025, Chen2025PRL}. The charge transfer leads to an upward positive vacuum level shift of the order of $ \Delta\phi=0.5$~eV. This was calculated from the \textit{local} variations of the work function when the molecule-probe height was $4$~\AA ~as shown in Fig.~\ref{fig:ips_and_eas}d.

After the charge transfer, the orbitals of the charged molecule re-order. Contrary to the non-transition metal-based molecules, here one needs to consider that the Fe center lies on top of the oxygen site of MgO, leading to electron-electron repulsion that shifts the $dz^2$ orbital upward (see SI section~\ref{sec:SI-fepc-1h2o} and Fig.~\ref{fig:fepc_orbital})~\cite{greule2025,Willke_2021}. This effect can be captured by placing a water molecule vertically below the Fe center of FePc$^{-1}$\cite{Urdaniz2025}. The molecular orbitals $\overline{\textrm{ORB}}_1$ and $\overline{\textrm{ORB}}_2$ show a character of $d_\pi$ and $d_{xy}$, respectively. Note that the $d_{z^{2}}$ orbital is partially filled, with one spin channel appearing close to the $\overline{\textrm{ORB}}_2$, while the other lies approximately 2~eV below the $\overline{\textrm{ORB}}_1$~\cite{greule2025}. After considering $\lambda$ and $P$ on the bare IP$=5.34$~eV and EA$=1.84$~eV of the FePc$^{-1}$, the $\overline{\textrm{ORB}}_1$ lies at $E'_{1}=-1.4$~eV an the $\overline{\textrm{ORB}}_2$ around $E'_{2}=0.9$~eV. Whilst the electronic gap $E_g'=2.3$~eV agrees well with the screened Hubbard $U$ from experiments ($U=2.6\pm0.6$~eV~\cite{Zhang_2025}) the charge extraction barrier $E_F-E'_{1} \equiv  \varepsilon\approx 1.4$~eV seems to underestimate experiments ($\varepsilon=-2\pm0.2$~eV~\cite{Zhang_2025}). This is most likely due to the \textit{local} potential variation which affects STS as it makes the vacuum level dependent on the probe-molecule distance, which was observed although not systematically analyzed in experiments~\cite{greule2025}. We note that the $\phi$ of the substrate is sensitive to preparation conditions which might alter stoichiometry or the interface structure~\cite{Nilius_2009, CHO2013541}. This might be an additional factor why charging energy $U$ is accurately described, whereas the barrier for charge extraction $\varepsilon$ is not.

\subsubsection{\texorpdfstring{$\pi$}{pi}-Conjugated Molecules}

\begin{table*}[!htb]
    \centering
    \small 
    \setlength{\tabcolsep}{4pt} 
    \begin{tabular*}{\textwidth}{@{\extracolsep{\fill}}c|ccc|cccc}
    \hline
        Adsorbate & IP~(eV)  &  EA~(eV)  & $E_g$~(eV) & $\Delta\phi$~(eV) & $\lambda$~(eV) & $E'_\mathrm{SOMO}$~(eV) & $E'_\mathrm{SUMO}$~(eV)\\
        \hline 
        \hline
        [FePc]$^{-1}$ & 5.34 & 1.84 & 3.5 & 0.50 & 0.05 & -1.4 & 0.9\\ 
        $[\textrm{Pentacene}]^{-1}$ & 4.41 & 0.80 & 3.61  & 0.52 & 0.10& -0.5 & 2.0 \\
        $[\textrm{PTCDA}]^{-1}$ & 5.66 & 2.32 & 3.34  & 0.59 & 0.10 & -0.9 & 1.6 \\
        $[\textrm{TCNE}]^{-1}$ & 7.21 & 1.68 & 5.53 & 2.24 & 0.09& -0.5 & 1.7 \\
        \hline
        \hline
        $[\textrm{Ti}]^{+1}$ & 5.06 & 3.05 & 2.01  & 0.71 & 0.29 & -1.4 & 0.6\\ \hline
    \end{tabular*}
    \caption{Summary of the physical quantities used in the energy alignment of adsorbates on oxide/metal substrates. IP and EA of the charged adsorbates [M]$^{q\pm}$ were calculated with the GW method. The local variation of the work function ($\Delta\phi$) was estimated in the GGA-PBE.  $\lambda$ is the reorganization energy. Values for $[\textrm{Ti}]^{+1}$ correspond to the frontier orbitals of the full system of Ti/MgO/Ag. Note that for Ti and FePc, the SOMO and SUMO correspond to the $\overline{\textrm{ORB}}_1$ and $\overline{\textrm{ORB}}_2$, respectively.}
    \label{tab:eaip3}
\end{table*}

\begin{figure}[!htb]
    \centering
    \includegraphics[width=0.80\linewidth]{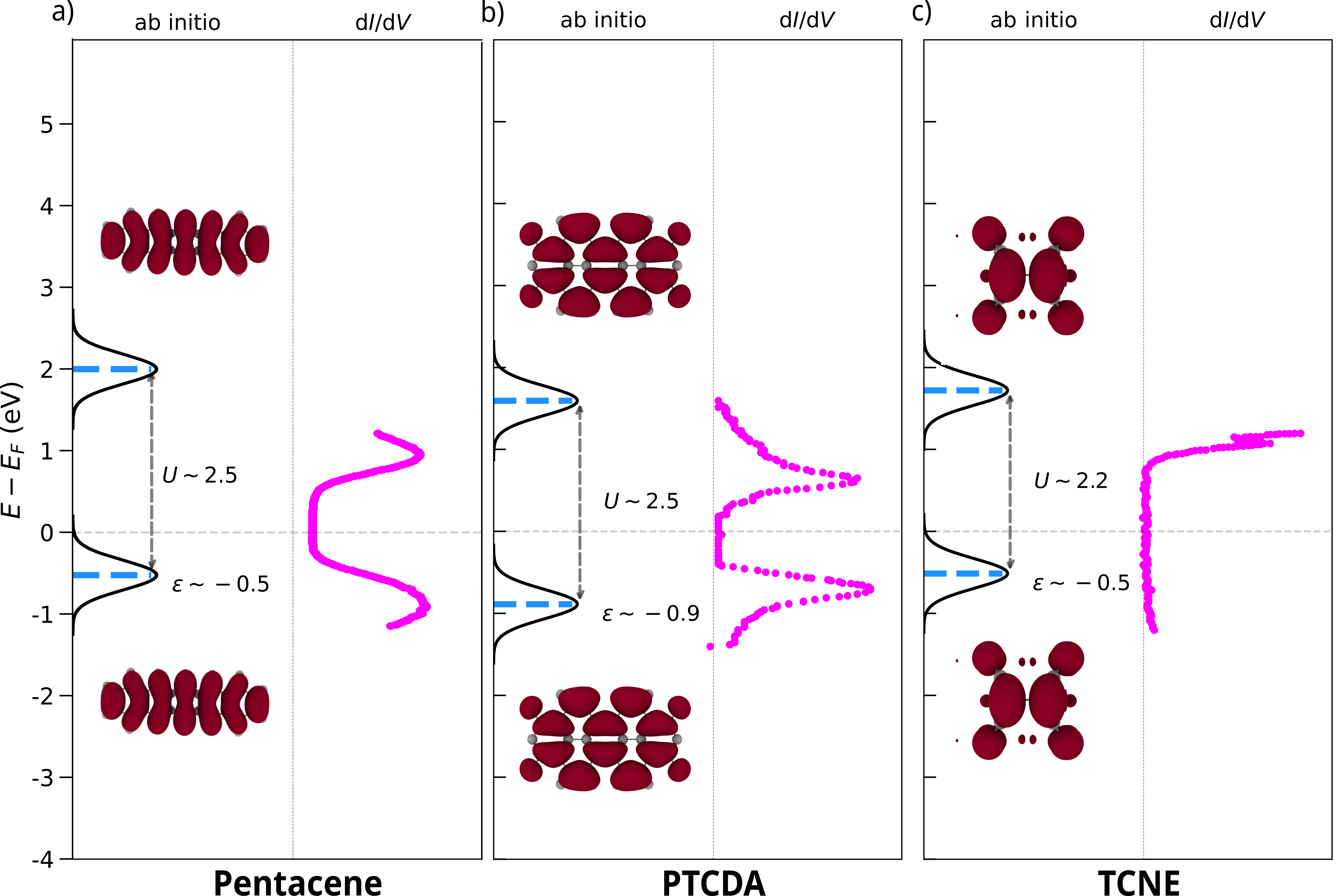}
    \caption{\textbf{Energy level alignment of Pentacene, PTCDA and TCNE} Shown are the frontier orbitals aligned relative to the respective substrate Fermi energy. (a) Pentacene/MgO/Ag, (b) TCNE/MgO/Ag, and (c) PTCDA/NaCl/Ag. Experimental STS data from Refs.~\cite{Hollerer2017,Zhang_2020,Cochrane_2015}. Charging energy and barrier for charge extraction are labeled as $U$ and $\varepsilon$. The orbital density of the frontier orbitals is shown as red isosurfaces (iso$=6.7\times10^{-3}$e/A$^3$).}
    \label{fig:pentacene_PTCDA_TCNE}
\end{figure}

Finally, we also test our predictions against three $\pi$-conjugated compounds: pentacene, PTCDA and TCNE. Pentacene (C$_{22}$H$_{14}$), is a ubiquitous molecule in surface science and organic electronics. Similarly, PTCDA is a molecule widely used for the study of molecular self-assembly on surfaces~\cite{Romaner_2009, Zhao2019}, as molecular single-photon emitter~\cite{Esat2018}, or as tip functionalization for ESR-STM~\cite{Esat2024}. TCNE is one of the strongest molecular electron acceptors owing to its four cyano groups, often forming anion or the doubly negatively charged dianion on metal surfaces~\cite{Erley1987, Hofmann2015}. The relevant ionization potentials, work function shifts and reorganization energies for these molecules are listed in Tab.~\ref{tab:eaip}. For detailed alignment diagram for each of these systems see supplemental material, Figs.~\ref{fig:pentacene}, \ref{fig:ptcda}, \ref{fig:tcne}. We find that these molecules form single anions. The resulting frontier orbitals aligned to the Fermi level of the respective substrates are shown in Fig.~\ref{fig:pentacene_PTCDA_TCNE}.

Pentacene has a relatively large electronic bandgap and a renormalized EA$'$ that is not immediately indicative of charge transfer. Experiments using photoemission, STM~\cite{Hollerer2017}, and ESR-STM~\cite{kovarik2024} have clearly shown that pentacene is charged on 2 ML of MgO deposited on Ag. The driving force behind the charge transfer is most likely a combination of the proximity of EA$'$ to $E_F$ and the reorganization of the molecule (see Fig.~\ref{fig:pentacene}). The best estimate from GW calculations brings the SOMO energy to $E'_\mathrm{SOMO}=-0.5$~eV and the SUMO energy to $E'_\mathrm{SUMO}=2$~eV, which is in good agreement with the experimental data. PTCDA on NaCl/Ag(111) exhibits an screened EA that leads to LUMO pinning and drives electron transfer from the substrate to the molecule. The resulting SOMO and SUMO energies of $-0.9$~eV and $1.6$~eV, respectively, are in good agreement with STM measurements reported in Ref.~\cite{Cochrane_2015}. This behavior is consistent with previous observations of negatively charged PTCDA on oxide–metal interfaces with comparable work functions, such as NaCl/Cu(111)~\cite{mohn_2010} and h-BN/Ni(111)~\cite{Schaal_2024}. A doubly charged regime is expected for substrates with work functions below 3.5~eV~\cite{Hurdax_2025}; however, NaCl/Ag(111) has a work function of 3.85~eV and no second charge transfer occurs. TCNE on MgO/Ag(001) show a comparable electron affinity to the work function of Ag/MgO substrate. This produces an alignment  that immediately places the LUMO close to the Fermi level, charging the molecule negatively. After considering the charge transfer, reorganization energy and polarization of the anion, the resulting frontier levels of the SOMO and SUMO energies of -0.5~eV and 1.7~eV, agree well with STS measurements~\cite{Wegner_2007,Zhang_2020} (see Fig.~\ref{fig:tcne}). The high electron affinity of TCNE indicates that it is a good model system for single-molecule ESR that should readily form an anion on most substrates.

\subsubsection{Ti atoms on MgO/Ag}
Lastly, we turn to single-atom adsorbates. Individual transition metal and lanthanide atoms have been studied using ESR-STM with the goal of building atomic-scale quantum bits. Among the 3$d$ transition metal series, particularly titanium (Ti) atoms on 2 and 3 ML of MgO have attracted interest due to their ability to form $S=1/2$ prototypical qubits,~\cite{Yang_2019, Wang2023science} and $S=1$ single atom magnets~\cite{Phark2026}. Recent experiments have concluded that Ti when adsorbed on 2 ML of MgO has a valence electron count of 3 (3$d^1$ 4$s^2$), which corresponds to a singly positively charged ([Ti]$^+$)~\cite{Phark2026}. ESR-STM experiments measured a renormalized transport gap of $U\approx 600$~meV and $\varepsilon\approx -250$~meV, see also Tab.~\ref{tab:eps_U})~\cite{Zhang_2025}. 

\begin{figure}[!htb]
    \centering
    \includegraphics[width=0.80\linewidth]{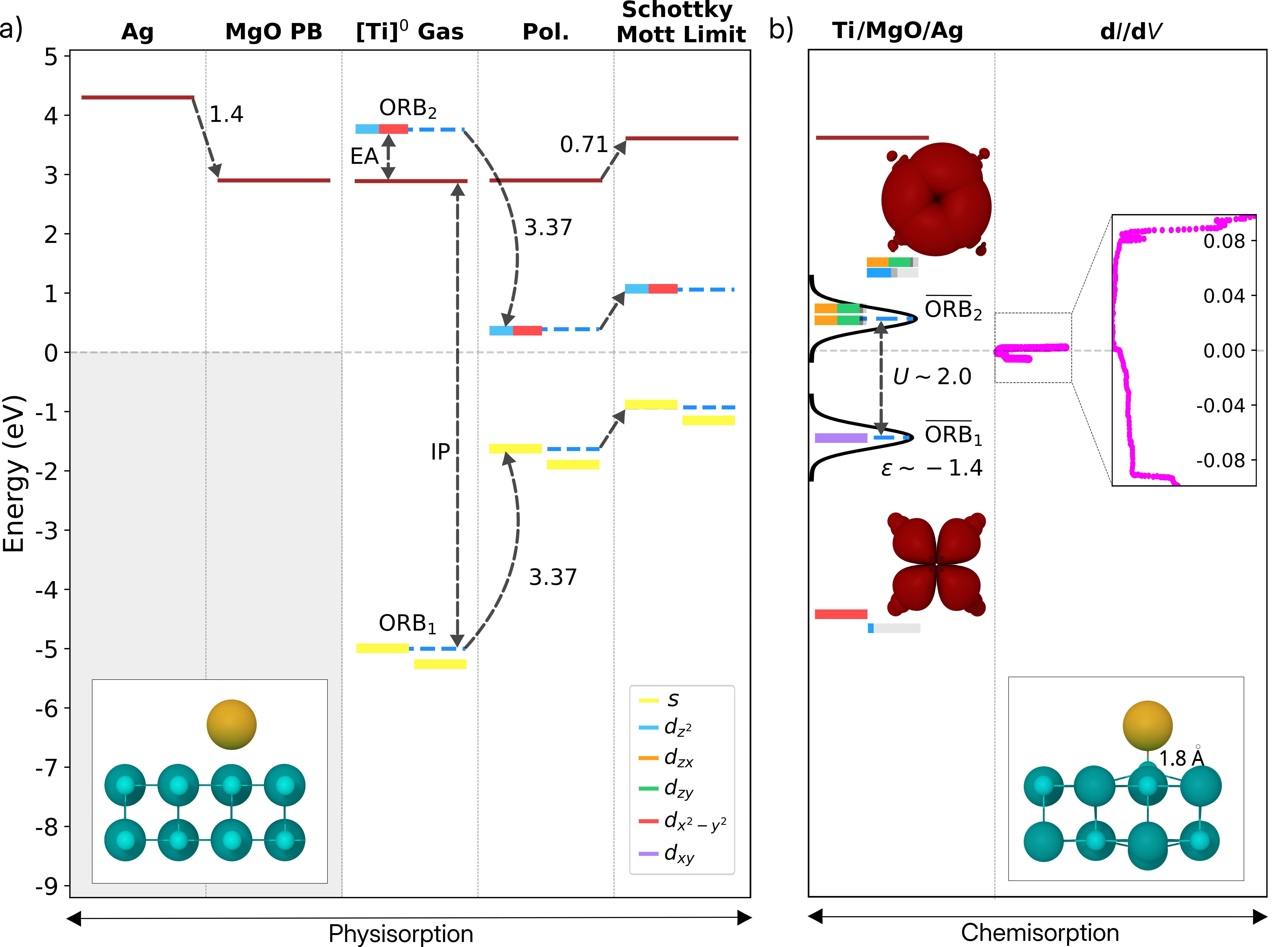}
    \caption{\textbf{Comparison between physisorption and chemisorption for Ti/MgO/Ag}. (a) The physisorption model shows the individual contributions leading to the energy alignment of Ti frontier orbitals. The inset shows the geometry of Ti adsorbed on the MgO surface in a physisorbed configuration. (b) The chemisorption is the simulation of the full Ti/MgO/Ag system (see inset for the geometry). Experimental STS data from Ref. \cite{Phark2026}.  The orbital density of the frontier orbitals is shown as red isosurfaces (iso=0.0005~$e/A^3$).}
    \label{fig:ti}
\end{figure}

We first apply the energy-level alignment rules for physisorbed systems to a Ti atom. Our calculations give a relatively high IP$\approx 7.9$ eV and a negative EA$\approx -0.9$ eV. The latter is most likely the consequence of DFT self-interaction error but experiments also indicate that Ti has a low EA which will not change the conclusions of the physisorption model. For consistency reasons, we proceed with the GW results in Fig.~\ref{fig:ti}a. Despite a strong renormalization ($2P=6.73$~eV), none of the frontier orbitals becomes pinned to the substrate Fermi level. As a consequence, the adsorbate stays neutral and the work function shift follows the Schottky–Mott limit. However, the 3$d^2$4$s^2$ electronic configuration of the physisorbed [Ti]$^0$ atom is incompatible with the ESR measurements, which detect a single-electron excitation~\cite{Phark2026,Yang_2017}. This suggests that describing single atoms on oxide/metal substrates requires an approach beyond the physisorption picture.

For this reason, we perform DFT calculations for the full Ti/MgO/Ag system. This calculation shows charge transfer from the adsorbate to the substrate, yielding [Ti]$^{+1}$, consistent with experiments~\cite{Phark2026}. This is the result of the inclusion of the chemical environment and covalent bonding of Ti on the MgO layer, further improved by the self-energy corrections captured by the GW method. The short 1.8~\AA ~Ti–O distance indicates hybridization between Ti orbitals and the oxide layer, consistent with chemisorption. Ti adsorbed on MgO most likely forms a cluster, with its IP and EA determined by the combined Ti–MgO electronic structure. Finally, our simulations of the complete Ti/MgO/Ag system are consistent with STS measurements (see Fig.~\ref{fig:ti}b), in the sense that they predict no ionic resonance within the experimentally accessible bias range of $V=\pm 200$~mV where experiments only show spin excitations, typically appearing at biases around $V=\pm 80\text{–}100$~mV~\cite{Phark2026}. However, the positions of the frontier orbitals are not consistent with the transport results of Ref.~\cite{Zhang_2025}, see Tab.~\ref{tab:eps_U}. To clarify the origin of this discrepancy, experiments revealing the orbital structure and ionic resonances of Ti on MgO will be required in the future.

\begin{table*}[!htb]
    \centering
    \small 
    \setlength{\tabcolsep}{4pt} 
    \begin{tabular*}{\textwidth}{@{\extracolsep{\fill}}l|ccccc} 
        \toprule
        Adsorbate & $E_g$~(eV) & Metal & Oxide & $E'_\mathrm{\overline{\mathrm{ORB}}_1}-E_F\equiv\varepsilon$~(eV) & $E'_g\equiv U$~(eV) \\
        \midrule \midrule
        \multirow{1}{*}{FePc}      & \multirow{1}{*}{3.33}  & Ag & MgO & -1.4 (-2.00~\cite{Zhang_2025}) & 2.16 (2.6~\cite{Zhang_2025}) \\
        \midrule
        Ti                         & 8.74                   & Ag & MgO & -1.4 (-0.25~\cite{Zhang_2025}) & 2.01 (0.6~\cite{Zhang_2025}) \\ 
        \bottomrule
    \end{tabular*}
    \caption{\textbf{Transport parameters of FePc and Ti on MgO/Ag.} The charge extraction barrier $E'_\mathrm{\overline{\mathrm{ORB}}_1}-E_F$ compared to $\varepsilon$ from experiments as well as $E'_g$ compared to $U$ from experiments.}
    \label{tab:eps_U}
\end{table*}

\section{Conclusions}
We have demonstrated a general and accurate fully ab initio based method to determine the charge state and frontier orbital alignment of molecular and atomic adsorbates. Our simulations show that whilst standard DFT methods are suitable to predict work functions of metals with ultrathin insulating layers they fail to predict reliable ionization potentials and electron affinities for atoms and molecules. To overcome the underestimation of the electronic bandgap we used the GW method. We showed that the COHSEX approximation for GW on top of GGA density is an excellent and computationally efficient method to obtain reliable energy level alignments for the molecules tested here. Most importantly, COHSEX GW@GGA accurately predicts substrate screening induced polarization $P$ of the IP and EA, which in turn critically influences the charge transfer of the adsorbate/substrate. A benchmark of our method against thoroughly experimentally studied systems pentacene/MgO/Ag, FePc/MgO/Ag, PTCDA/NaCl/Ag, and TCNE/MgO/Ag, highlights the strength of a fully ab initio-based method. This method can be used for ab initio quantum materials discovery by designing adsorbate-substrate combinations with desired charge and spin states. Finally, for the case of Ti/MgO/Ag, the results indicate that the system cannot be described by our physisorption method, suggesting that single atoms on oxide/substrate systems require the simulation of the full system.


\section{Methods}
\subsection{Plane-wave pseudopotential DFT calculations}
Ab initio methods were performed using density functional theory with pseudopotentials and  plane-wave basis implemented in Quantum Espresso V~7.5. We used norm-conserving pseudopotentials from the ONCV library for all ions~\cite{Hamann2013}. The cutoff for the expansion of the basis in plane-waves was set to $E_\mathrm{cut}$=90 Ry. Our DFT calculations used the generalized gradient approximation (GGA-PBE)~\cite{Perdew1996} except for Ti atoms, where we used B3LYP as functional to generate the ground state density. For non self-consistent calculations we set the number of empty bands to 8 times the number of filled states in all our calculations. For calculations of adsorbates in the gas phase we used the "particle-in-a-box" by placing the atom or molecule in a vacuum box of length $16$~\AA  ~ for molecules like benzene and TCNE, while larger molecules like pentacene, PTCDA and FePc were placed in boxes with dimensions of $13\times24\times15$~\AA$^3$, $18\times23\times14$~\AA$^3$ and $20\times20\times15$~\AA$^3$, respectively, to isolate them from periodic images.

For the calculation of the potential variations that require the full system simulations, the unit cell of the substrate of NaCl/Cu(001) (with $a=5.71$ ~\AA) was replicated by $2\times2\times1$  for benzene/NaCl/Cu(001). Similarly, the unit cell of the substrate of MgO/Ag(001) (with $a=4.16$~\AA) was replicated by $6\times3\times1$ for pentacene/MgO/Ag(001), $5\times5\times1$ for FePc/MgO/Ag(001) and $3\times3\times1$ for TCNE/MgO/Ag(001). Lastly, the unit cell of the substrate of NaCl/Ag(111) (with $a=8.83$~\AA ~and $b=10.61$~\AA ~ with an angle of 74$^\circ$) was replicated by $3\times2\times1$ for PTCDA/NaCl/Ag(111). Note that all substrates contain 2 layers of oxide and 2 layers of metal. Additionally, for the adsorbate/oxide system used in the polarization calculations, we employed the same supercell as in the full adsorbate/oxide/metal system, but with the metal layer removed. This ensures consistency in geometry and periodicity while isolating the oxide contribution. All calculations were performed using only the $\Gamma$-point sampling of the Brillouin zone. For periodic slab calculations we generated suitable lateral supercells of the substrate. Surface slabs were decoupled in $z$-direction by using the dipole method with the dipole placed $~10$ ~\AA ~above the molecule. We note that this is critical for charged cells due to the long-ranged Coulomb interactions.

\subsection{GW calculations}
GW calculations were performed based on the DFT ground state density using YAMBO (V 5.0.3)\cite{Sangalli_2019, MARINI20091392}. In molecules, dielectric screening is weak and dominated by high-energy electronic excitations, so the frequency dependence of the screened interaction $W(r,r',\omega)$ is relatively small near the frontier orbital energies. As a result, the static interaction $W(r,r',\omega=0)$ captures the dominant polarization response associated with the formation of the Coulomb hole, yielding reliable level shifts while avoiding the computational cost of a fully dynamical treatment~\cite{hedin1965,Aryasetiawan_1998}. 

For calculations of Ti, we performed GW calculations using both GGA and B3LYP starting densities. For the isolated system, the ionization potential (IP) and electron affinity (EA) were computed on top of the B3LYP solution, which yields a larger electronic bandgap and therefore provides a more reliable description of the HOMO and LUMO levels. The choice of starting point is important in $GW$ calculations because the results retain a dependence on the initial electronic structure: functionals such as GGA typically underestimate the electronic bandgap, leading to overscreening and reduced quasiparticle corrections, whereas B3LYP, with its larger electronic bandgap, generally reduces overscreening and improves the predicted frontier orbital energies  for single isolated atoms~\cite{Golze_2019,rostgaard2010}.

\section{Acknowledgments}
The authors acknowledge support from the Institute for Basic Science under grant IBS-R027-D1. We acknowledge fruitful discussions with Robert Ranecki, Franklin H. Cho, Yaowu Liu, Xue Zhang, Xiaobin Geng.

\clearpage
\onecolumngrid 

\begin{center}
    {\Large \textbf{Supplementary Material for:}}\\[0.5em] 
    {\normalsize \textbf{An ab initio approach to energy alignment and charge-state prediction of adsorbates on ultrathin insulators}} \\[1.5em]
    
    Kevin Lizárraga$^{1,2}$, Saba Taherpour$^{1,3}$, Cesar E. P. Villegas$^{4}$, and Christoph Wolf$^{1,2}$\\[1em]
    
    \textit{$^{1}$Center for Quantum Nanoscience, Institute for Basic Science (IBS), Seoul 03760, Republic of Korea\\
    $^{2}$Ewha Womans University, Seoul 03760, Republic of Korea\\
    $^{3}$Department of Physics, Ewha Womans University, Seoul 03760, Republic of Korea\\
    $^{4}$Departamento de Ciencias, Universidad Privada del Norte, Lima 15434, Peru.}
\end{center}

\vspace{2em} 

\appendix 
\makeatletter 

\setcounter{section}{0}
\setcounter{figure}{0}
\setcounter{table}{0}
\setcounter{equation}{0}

\renewcommand{\thesection}{S\arabic{section}} 
\renewcommand{\thefigure}{S\arabic{figure}}
\renewcommand{\thetable}{S\arabic{table}}
\renewcommand{\theequation}{S\arabic{equation}}

\renewcommand{\p@section}{}

\makeatother














\section{Details of the GW calculations}
\subsection{Notes on GW parameters}

We have performed converged calculations within the framework of the GW COHSEX approximation for the isolated molecules. For non–transition-metal adsorbates such as benzene, pentacene, PTCDA, and TCNE, convergence was achieved using a cutoff for the exchange–correlation potential matrix elements ($\textrm{VXCRLvcs}$) of 300000~RL, a cutoff for the exact exchange part of the self-energy ($\mathrm{EXXRLvcs}$) of 60~Ry, and a cutoff for the screened Coulomb interaction ($\mathrm{NGsBlkXs}$) of 10~Ry. The number of bands included for the polarization was set to eight times the number of filled bands. For transition-metal adsorbates such as FePc and Ti. We employed $\textrm{VXCRLvcs} = 900000$~RL, $\mathrm{EXXRLvcs}= 60$~Ry, and $\mathrm{NGsBlkXs}= 12$~Ry, and the number of polarized bands was set to 8 times the number of valence bands. Lastly, for the calculations on the substrates, we have used the same parameters as in the case of isolated molecules.

\section{Impact of the metal on the polarization of Benzene}
\label{sec:SI-benzene-pol}
\begin{figure*}[!htb]
    \centering\includegraphics[width=0.85\linewidth]{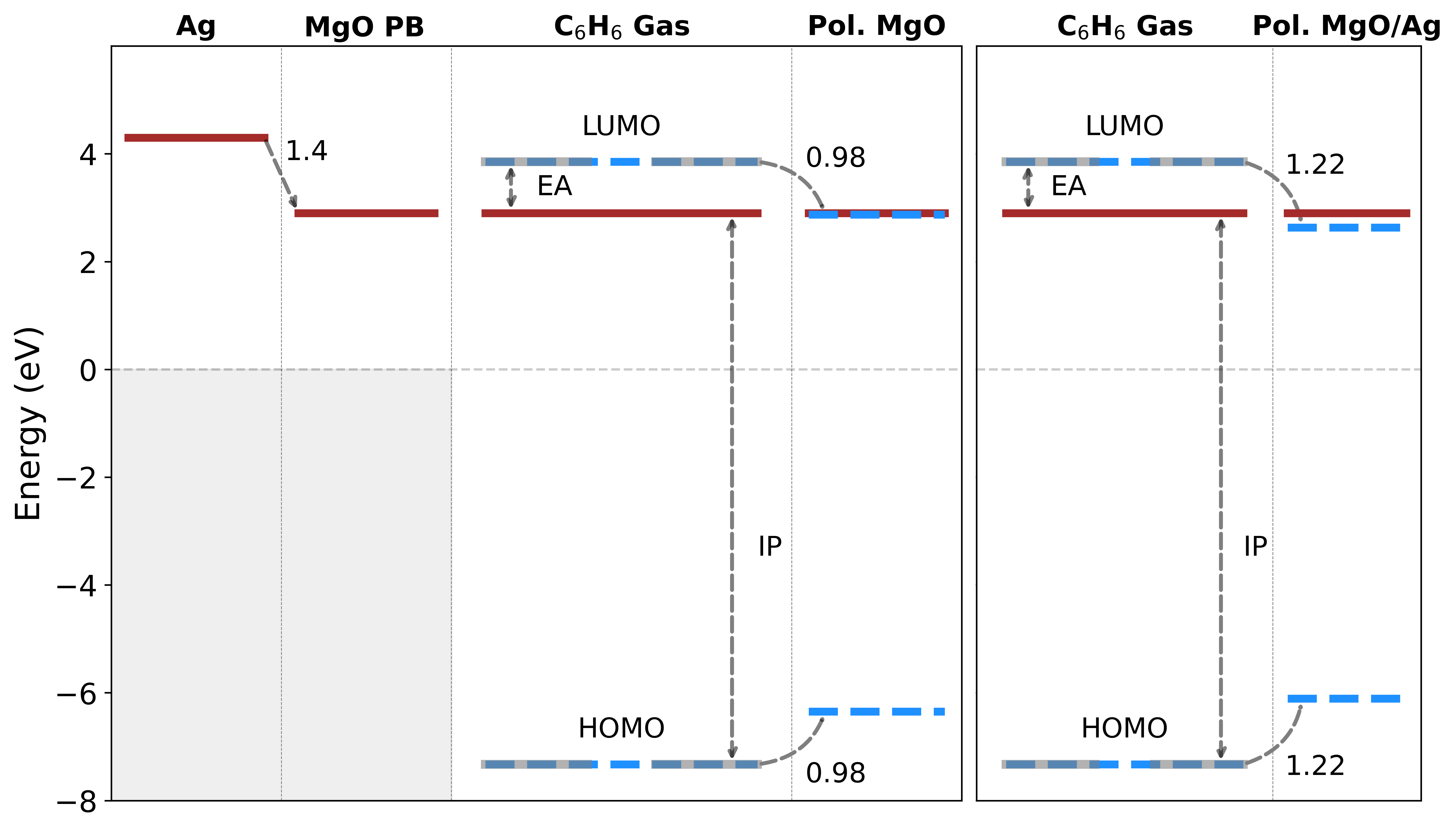}
    \caption{Left: Polarization of benzene when deposited on top of MgO only. Right: benzene when deposited on top of MgO/Ag. The presence of the metal substrate increases the screening.}
    \label{fig:pol_benzene}
\end{figure*}

Figure \ref{fig:pol_benzene} shows the effect of MgO and MgO/Ag substrates on the polarization of benzene. The presence of the Ag support increases the polarization by approximately 0.5~eV compared to MgO alone. This additional polarization originates from the enhanced screening provided by the metallic substrate beneath the insulating MgO layer. For the FePc/MgO/Ag, pentacene/MgO/Ag, and PTCDA/NaCl/Ag systems, explicit GW calculations including the Ag substrate exceeded our computational resources. To account for the metallic support, we therefore added an additional $2\tilde{P} = 0.5$ eV to the polarization obtained for the FePc/MgO, pentacene/MgO, and PTCDA/NaCl systems. This approach allows us to effectively capture the screening effect of the Ag substrate while avoiding the computational cost of full GW simulations.

\subsection{Comparison of GW calculations and \texorpdfstring{$\Delta$}{Delta-} SCF calculations}
\begin{figure*}[!htb]
    \centering\includegraphics[width=1.0\linewidth]{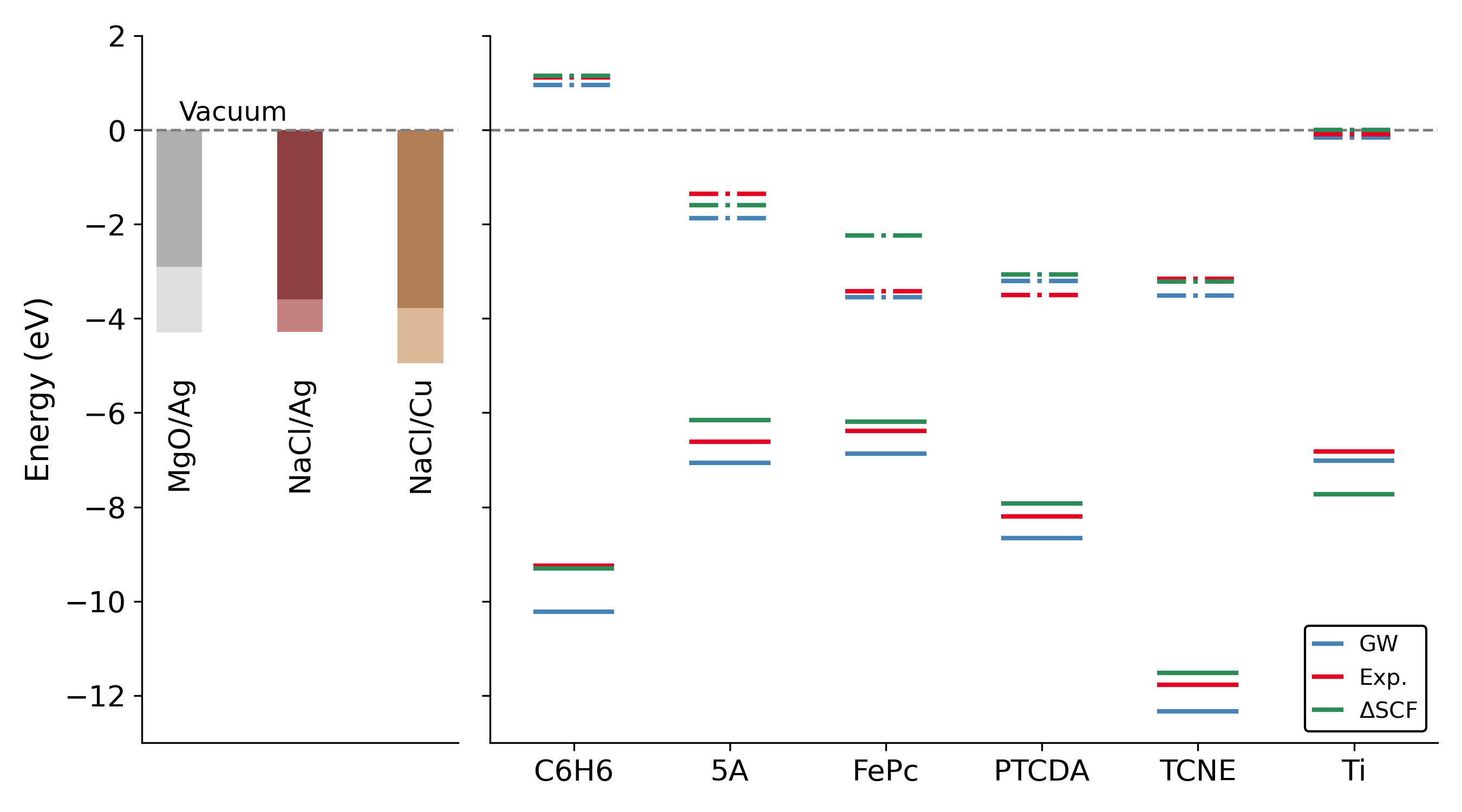}
    \caption{Right panel shows the work function $\phi$ of the oxide/metal substrates. Left panel shows the comparison of the IP and EA for the adsorbates calculated with GW and $\Delta$SCF method along with the experimental results.}
    \label{fig:supp_deltascf_GW}
\end{figure*}

Fundamental electronic gaps of atoms and molecules can also be calculated by the so-called $\Delta$SCF method, which uses the total energies of SCF calculations with different charge states to estimate the energy of removing (adding) one electron, i.e. the EA and IP. Figure \ref{fig:supp_deltascf_GW} shows the comparison of the $\Delta$SCF method with the GW method. All values for the adsorbate compare well, except for the case of FePc and Ti. The $\Delta$SCF values were obtained using the TPSSh functional, D3BJ van der Waals correction and a def2-TZVP basis as implemented in ORCA\cite{Neese_2022,Garcia_Rat_s_2019}. The unfavorable scaling of this method which system size limits its application to small isolated molecules~\cite{Neese_2022}.

\section{Local vacuum level shift from interface dipole formation}
\label{sec:SI-capacitor}

The work function of the metal is first altered by the Pauli pushback of the oxide $\Delta\phi_\textrm{PB}=\phi_\textrm{metal}-\phi_\textrm{oxide}$ (see also Fig.~\ref{fig:ips_and_eas}(d) in the main text). Then, the resulting work function can be further altered by the adsorbate, which we calculated as:
\begin{equation}
    \Delta\phi=\left( \Delta\phi_\textrm{PB} - (V_\textrm{metal}-V_\textrm{adsorbate})  \right).
\end{equation}

\begin{figure*}[!htb]
    \centering\includegraphics[width=0.7\linewidth]{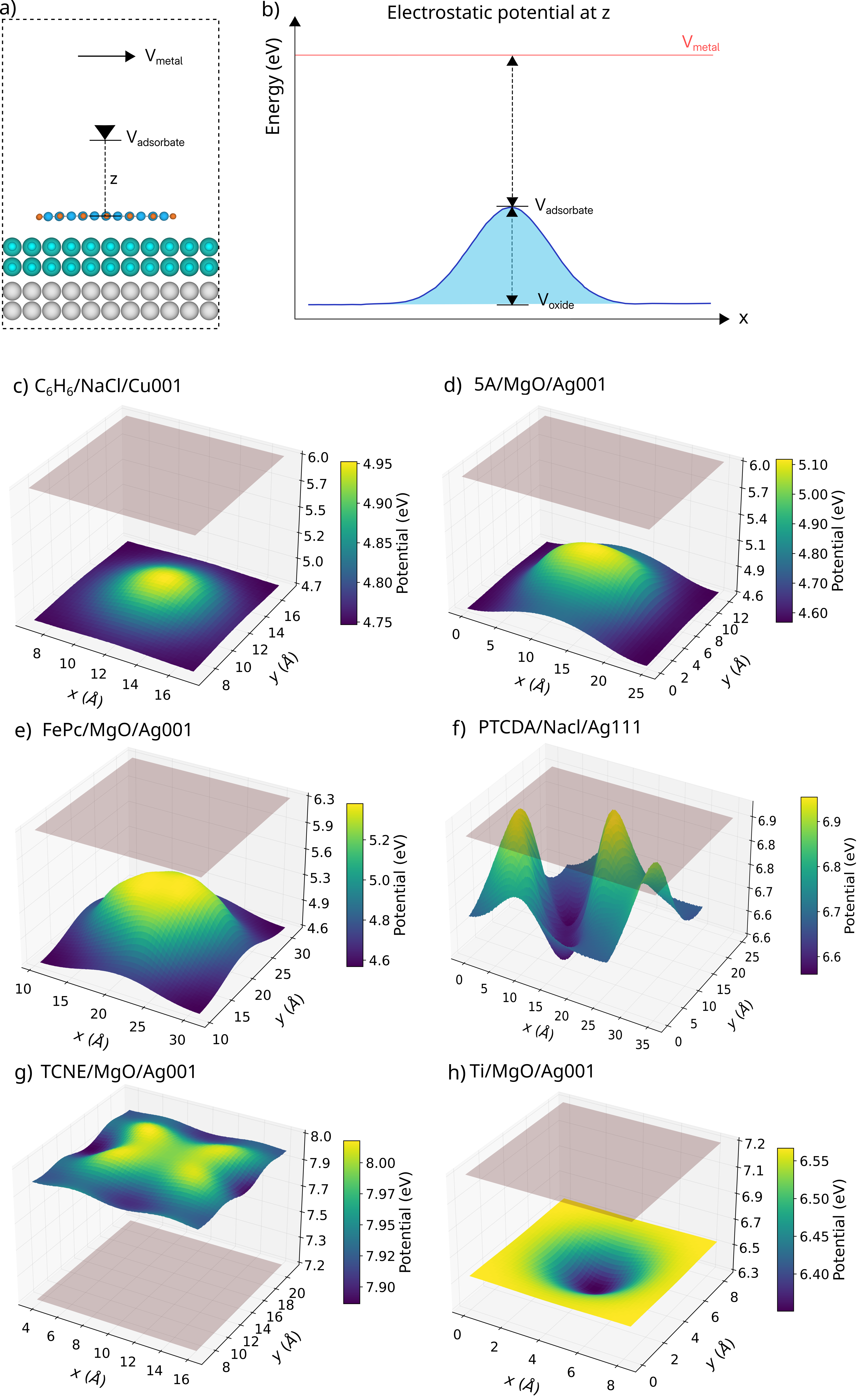}
    \caption{(a) Schematic of a scanning-probe setup that probes the electrostatic potentials of the metal ($V_\textrm{metal}$) and the adsorbate ($V_\textrm{adsorbate}$). (b) Schematic of the electrostatic potential at given height $z$. Local potential variation for the systems of (c) C$_6$H$_6$/NaCl/Cu(001) (d) pentacene/MgO/Ag(001), (e) FePc/MgO/Ag(001), (f) PTCDA/NaCl/Ag(111), (g) TCNE/MgO/Ag(001) and (h) Ti/MgO/Ag(001). For all plots, we have chosen an adsorbate-probe height, $z=4$~\AA ~and the horizontal plane indicates the work function of the metal for each plot.}
    \label{fig:charge_transfer}
\end{figure*}

\begin{figure*}[!htb]
    \centering\includegraphics[width=0.75\linewidth]{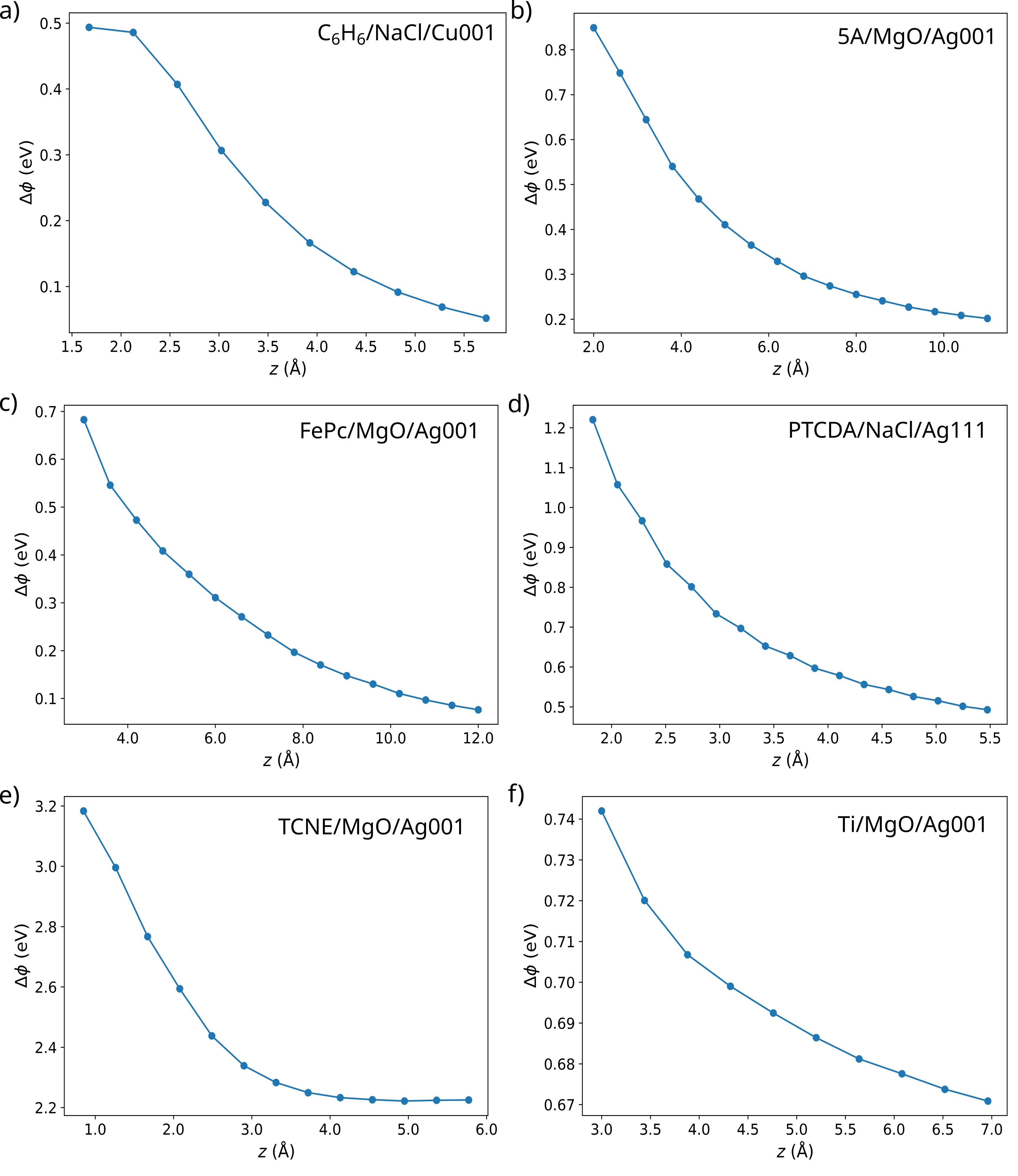}
    \caption{Local potential variation as function of $z$ of the probe to the adsorbate for (a) C$_6$H$_6$/NaCl/Cu(001) (b) pentacene/MgO/Ag(001), (c) FePc/MgO/Ag(001), (d) PTCDA/NaCl/Ag(111), (e) TCNE/MgO/Ag(001) and (f) Ti/MgO/Ag(001). The dashed gray lines show the value used in this work.}
    \label{fig:WF_height}
\end{figure*}

The electrostatic difference $(V_\textrm{metal}-V_\textrm{adsorbate})$ is obtained from simulations of the full adsorbate/oxide/metal system, which is necessary to capture the dipole formed due to the charge transfer. Here, $V_\textrm{metal}$ is taken from the electrostatic potential in the metal region, which is decoupled from the molecule by the artificial dipole correction (see Fig.~\ref{fig:charge_transfer}). The adsorbate-induced $V_\textrm{adsorbate}$ is defined as the maximum electrostatic potential at a given height $z$ (see Fig.~\ref{fig:charge_transfer}) from the adsorbate. $\Delta\phi$ strongly depends on $z$, see Fig.~\ref{fig:WF_height}. For local STM or STS measurement  $z\sim4$~\AA ~show $\Delta\phi\sim1$~eV, in agreement with the STM measurements on PTCDA and Pentacene~\cite{Hollerer2017,Schaal_2024}. In contrast, at larger distances ($z\ge 10$~\AA), $\Delta\phi$ is significantly reduced. This correspond to techniques that probe larger surface areas and effectively average over extended regions such as ultraviolet photoelectron spectroscopy (UPS). The values of $\Delta\phi \approx 0.15$~eV are in good agreement with Refs.~\cite{Greiner_2013,Ley_2012,Chai_2014}

It is important to note that $\Delta\phi$ cannot be directly determined from $V_\textrm{oxide}-V_\textrm{adsorbate}$ (see Fig.~\ref{fig:charge_transfer}(b)), because the potential from the oxide, $V_\textrm{oxide}$, varies with $z$. In contrast, we use the $V_{\textrm{metal}}$ because it is constant.

The values of $\Delta\phi$ at $z = 4$~\AA ~(gray lines in Fig.~\ref{fig:charge_transfer} and listed in Table~\ref{tab:work_functions}) reflect the magnitude of the interface dipole. For C$_6$H$_6$ the negligible charge transfer leads to only minor variations in $\Delta\phi\approx 0.2$. In contrast, for singly charged FePc/MgO/Ag, pentacene/MgO/Ag, PTCDA/NaCl/Ag, the work-function shifts positively by $\Delta\phi\approx 0.5$. For TCNE/MgO/Ag, the large positive shift of $\Delta\phi=2.24$~eV comes from the small spatial localization of the charge in the molecule. In the simulation of the Ti/MgO/Ag system, we find $\Delta\phi=0.71$~eV, even though charge is transferred from the adsorbate to the substrate. Unlike physisorbed systems, where the sign of $\Delta\phi$ directly reflects the direction of charge flow, the chemisorption of Ti induces a localized charge at the Ti–O bond, resulting in an interface dipole that raises the work function.

\begin{table*}[!htb]
    \centering
    \small 
    \setlength{\tabcolsep}{4pt} 
    \begin{tabular*}{\textwidth}{@{\extracolsep{\fill}}c|ccc|cccc}
    \hline
        System & $\Delta\phi_\textrm{PB}$~(eV)  &  $V_\textrm{metal}$~(eV)  & $V_\textrm{adsorbate}$~(eV) & $\Delta\phi$~(eV)\\
        \hline 
        \hline
        C$_6$H$_6$/NaCl/Cu(001) & 1.20 & 5.84 & 4.92 & 0.28\\ 
        FePc/MgO/Ag(001) & 1.40 & 6.28 & 5.38 & 0.50\\ 
        pentacene/MgO/Ag(001) & 1.40 & 6.00 & 5.12 & 0.52\\
        PTCDA/NaCl/Ag(111) & 0.55 & 6.92 & 6.96 & 0.59\\
        TCNE/MgO/Ag(001) & 1.40  & 7.17 & 8.01 & 2.24\\
        \hline
        \hline
        Ti/MgO/Ag(001) & 1.40  & 7.25 & 6.56 & 0.71\\ \hline
    \end{tabular*}
    \caption{Parameters used for the calculation of the work function shift $\Delta\phi=\left( \Delta\phi_\textrm{PB} - (V_\textrm{metal}-V_\textrm{adsorbate})  \right)$. These include the Pauli pushback ($\Delta\phi_\textrm{PB}$), the electrostatic potential of the metal ($V_\textrm{metal}$) and the adsorbate ($V_\textrm{adsorbate}$).}
    \label{tab:work_functions}
\end{table*}

\section{Comparison between molecular orbitals of FePc/MgO/Ag and \texorpdfstring{[FePc]$^{-1}$}{[FePc]^-1} on top of \texorpdfstring{H$_2$O}{H2O}}
\label{sec:SI-fepc-1h2o}

\begin{figure*}[!htb]
    \centering\includegraphics[width=0.85\linewidth]{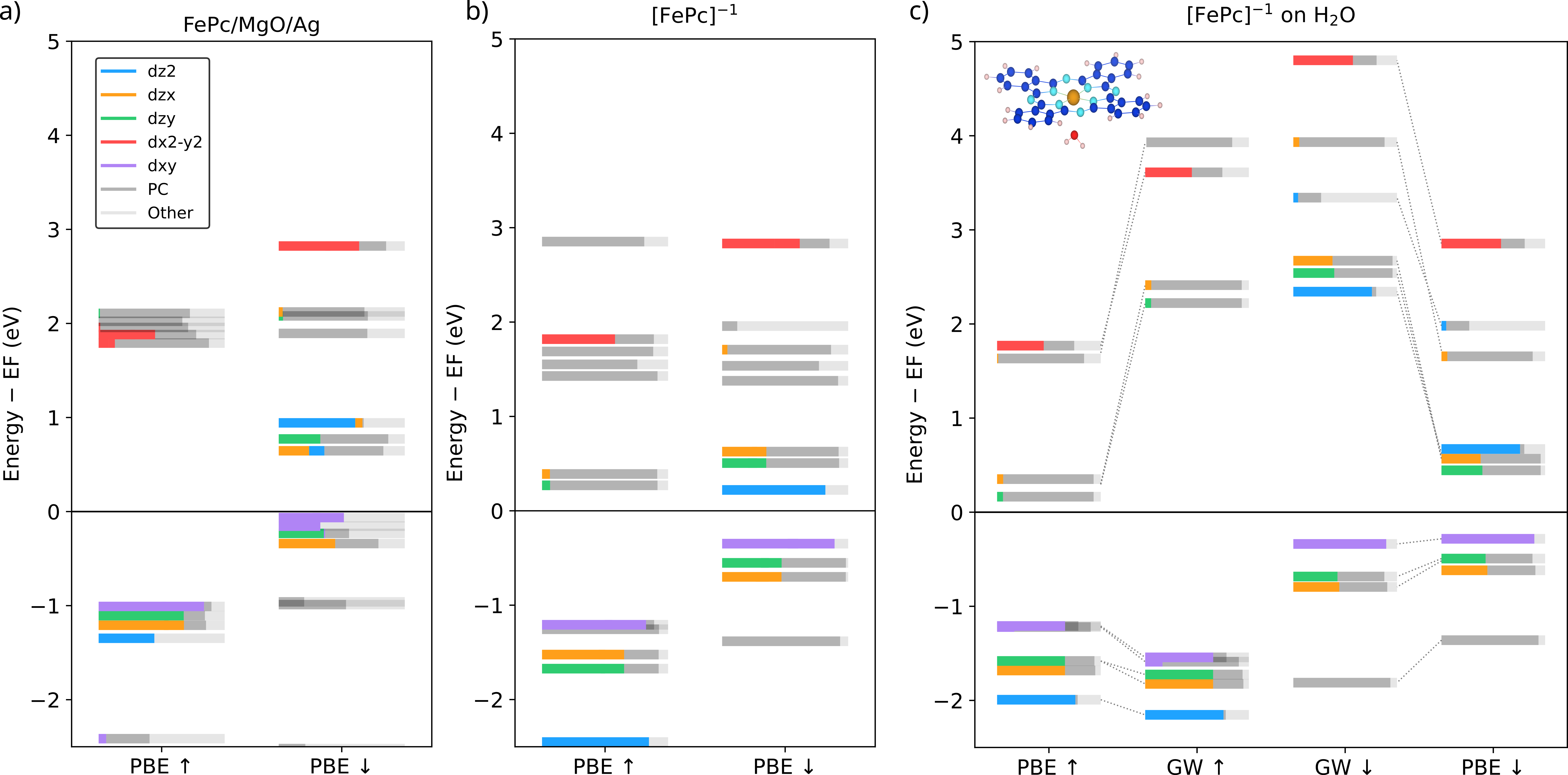}
    \caption{(a) Molecular orbitals of the FePc in the full organic-substrate system FePc/MgO/Ag. (b) Molecular orbitals of the isolated [FePc]$^{-1}$. (c) Molecular orbitals of the [FePc]$^{-1}$ on top of H$_2$O where oxygen points toward the Fe in the configuration (see inset) in the GW and PBE levels. The GW orbitals are the ones shown in Fig.~\ref{fig:fepc} of the main text.}
    \label{fig:fepc_orbital}
\end{figure*}

When simulating the isolated charged molecule of FePc in the PBE level, we found that the orbital ordering when compared to the FePc/MgO/Ag system of the $d\pi$ and $dz^2$ states are inverted in the down channel (see Fig.~\ref{fig:fepc_orbital}a and b)This inversion arises from the additional Fe–O bonding present in the full organic–substrate system. Consequently, the isolated [FePc]$^{-1}$ configuration does not accurately reproduce the molecular orbital ordering of the complete interface. To achieve the same molecular level ordering at the PBE level we placed a single water molecule below the Fe atom\cite{Urdaniz2025}, which results in an orbital order that closely resembles that of the full organic–substrate system, see Fig.~\ref{fig:fepc_orbital}c. Such an improved initial guess is particularly important for the GW calculations, which reproduce the near degeneracy between the $d_\pi$ and $d_{z^2}$ molecular orbitals.

\section{Detailed energy alignment plots}
\subsection{Detailed alignment plots for pentacene}\label{sec:SI-pentacene}

\begin{figure*}
    \centering
    \includegraphics[width=1.00\linewidth]{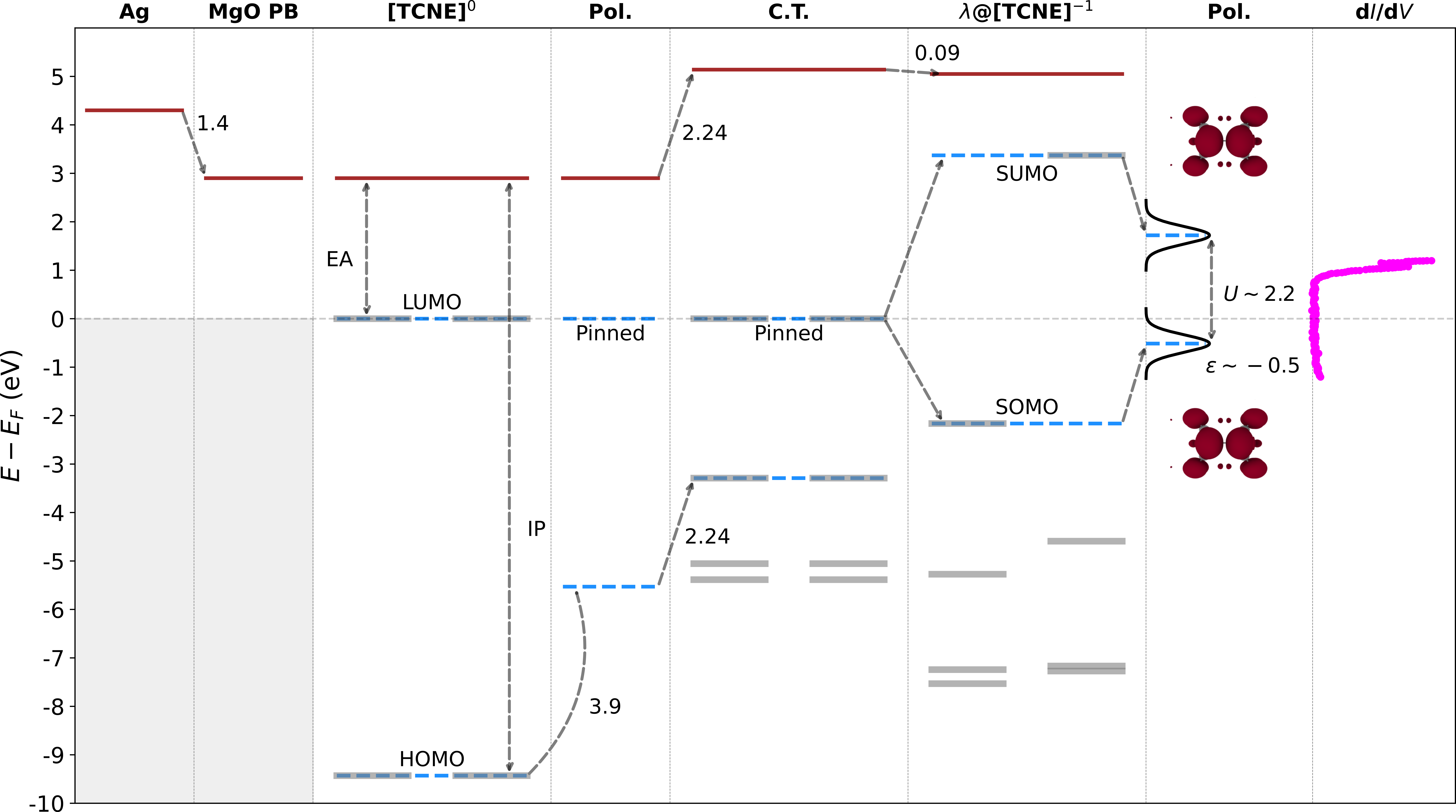}
    \caption{\textbf{Energy level alignment of pentacene on Ag/MgO.} The schematic from left to right correspond to all the effects such as Pauli pushback ($\Delta\phi_\textrm{PB}$), polarization (Pol.), charge transfer (C.T.), and reorganization of energy levels. Molecular orbital density plots of the SOMO and SUMO appear as red isosurfaces. Experimental data from Ref.\cite{kovarik2024}. }
    \label{fig:pentacene}
\end{figure*}

For pentacene deposited on MgO/Ag, we align the vacuum level to the work function of the interface Ag(001)  $\phi=4.3$~eV and two monolayers of MgO with a Pauli pushback of $\Delta \phi_\textrm{PB}=1.4$~eV. The alignment of the HOMO and LUMO levels (where the IP$=6.61$~eV and EA$=1.35$~eV), produces a LUMO energy level located at $\sim1$~eV above the Fermi energy. However, this is affected by the polarization of the electronic bandgap of $2P= 1.1$~eV, resulting in renormalized IP and EA of $0.55$~eV and $0.55$~eV, respectively, as can be seen in Fig.~\ref{fig:pentacene}. Despite the lack of clear Fermi level pinning behavior, the proximity of the LUMO level to the Fermi energy, enables a charge transfer from the metal substrate~\cite{Ley_2012,Chai_2014}, after which the LUMO will be pinned. The shift in the vacuum level after charge transfer is $\Delta \phi=0.52$ eV, which is comparable to previous estimates of $\sim1$~eV~\cite{Hollerer2017}. Note that, differences can be attributed to the $locality$ of the STM measurement. In our simulations, $\Delta \phi$ was calculated when the probe-molecule height was $4$~\AA. Nonetheless, values of $\sim0.8$~eV can be obtained at lower distances of $2$~\AA (see Fig. \ref{fig:WF_height}(c)).

The GW calculation of the anion pentacene$^{-1}$ and the inner shell reorganization energy $\lambda=0.1$ eV, lead to the energy alignment shown in the seventh column with $E'_\mathrm{SOMO}= -0.5$, $E'_\mathrm{SUMO}= 2.0$~eV, and $E_g'=2.5$~eV. This leads to values of Hubbard and charging energy of $U=2.5$~eV and $\varepsilon=-0.5$~eV, respectively, which agrees well with  the results of the d$I$/d$V$ spectroscopy of singly charged $S=1/2$ on MgO/Ag~\cite{Hollerer2017, kovarik2024}. It is important to note that STS does not measure the \textit{Kohn-Sham orbital} but rather reveals the transition between orbitals belonging to different charge states of a molecule, i.e. the positive and negative ionic resonance (PIR, NIR). An accurate description of the orbital character therefore involves Dyson orbitals~\cite{kumar2025theoryscanningtunnelingspectroscopy}, however the orbital character is beyond the scope of this work where we are mostly concerned with the energy position of the frontier orbitals relative to the vacuum level. In Ref. \cite{Hollerer2017} a similar orbital level alignment was deduced from a combination of ab initio simulations and experimental measurements. In contrast, our approach is fully ab initio.

\subsection{Detailed alignment plots for PTCDA}\label{sec:SI-ptcda}

\begin{figure*}
    \centering
    \includegraphics[width=0.85\linewidth]{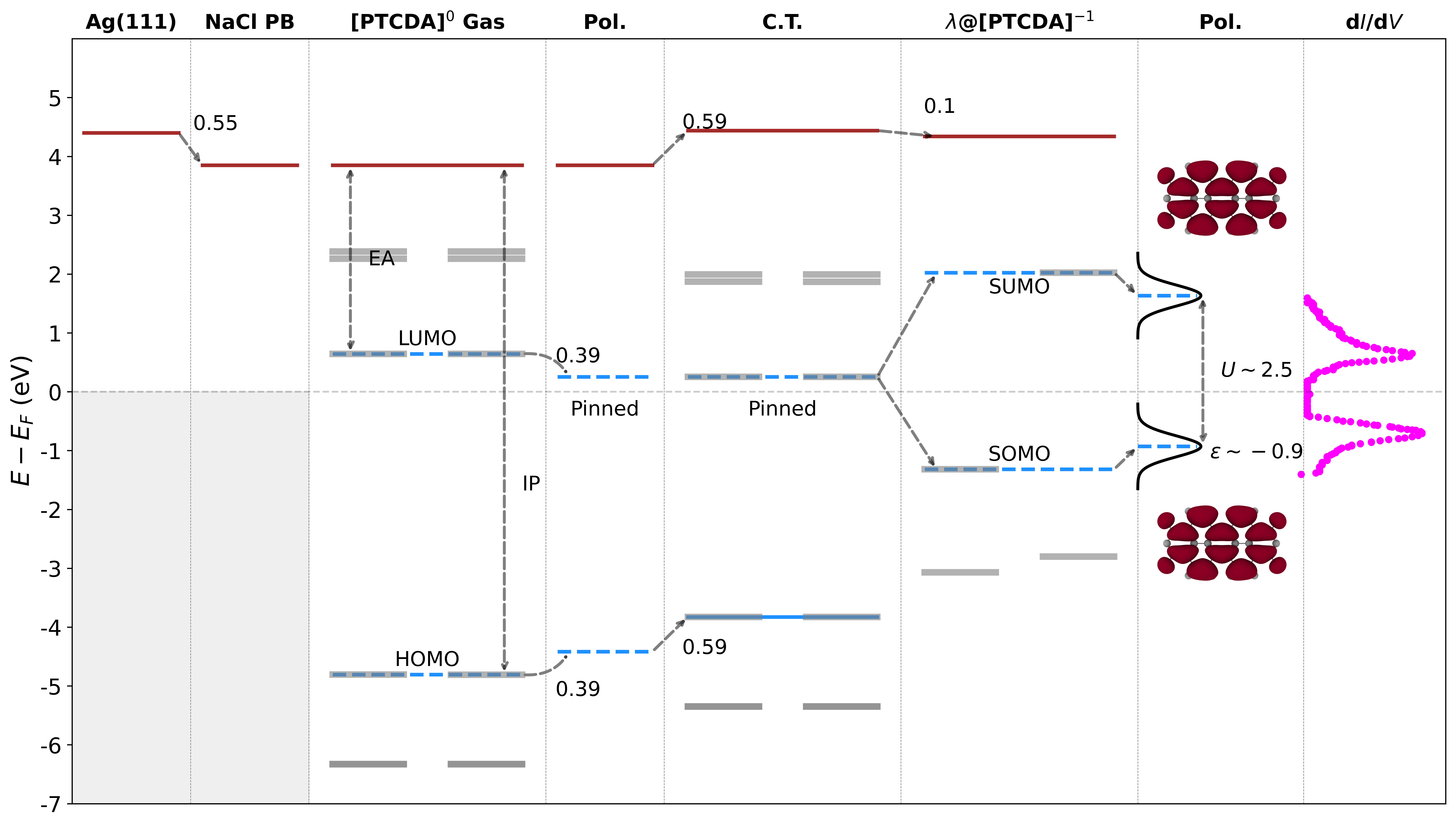}
    \caption{\textbf{Energy level alignment of PTCDA on Ag/NaCl.} The schematic from left to right correspond to all the effects such as Pauli pushback ($\Delta\phi_\textrm{PB}$), polarization (Pol.), charge transfer (C.T.), and reorganization of energy levels. In the experiment column, the data of [ref] is shown in magenta dots, the Gaussians correspond to our prediction, and the molecular orbitals of the SOMO and SUMO appear as red isosurfaces.}
    \label{fig:ptcda}
\end{figure*}

Pristine PTCDA (C$_{24}$H$_8$O$_6$) is a closed shell system with  IP$=8.65$~eV and EA$=3.20$~eV and a electronic bandgap of $5.05$~eV  \cite{kang2016}. The alignment with respect to the vacuum level of the NaCl/Ag(111), located $3.85$~eV above the Fermi energy--after the PB effect of $\Delta\phi_\textrm{PB}=0.55$--, does not initially indicate charge transfer upon adsorption. Substrate screening reduces the HOMO–LUMO gap by $2P=0.78$~eV. The proximity of the LUMO level to the Fermi energy enables a charge transfer and the pinning of the LUMO (see Fig.~\ref{fig:ptcda}). This in agreement with experimental measurements~\cite{Cochrane_2015}. We obtained an upward shift of the vacuum level of approximately $\Delta\phi=0.59$~eV (see SI Section \ref{sec:SI-capacitor}). this value is comparable to the work function change of $\sim0.7$~eV measured in I(z) spectroscopy for PTCDA/h-BN/Ni(111) system~\cite{Schaal_2024}. Once charge transfer takes place, the [PTCDA]$^{-1}$ has an IP$=5.66$~eV and EA$=2.32$~eV (see the sixth column of the Fig. \ref{fig:ptcda}). By accounting for the inner relaxation energy ($\lambda$) and the polarizability, we obtain SUMO and SOMO orbitals at $E'_\mathrm{SOMO}= -0.9$ and $E'_\mathrm{SUMO}= 2.5$~eV, respectively, along with an electronic bandgap of $2.0$~eV. These leads to a Hubbard repulsion of $U = 2.5$~eV and a charging energy of $\varepsilon = -0.9$~eV. Our results are in agreement with d$I$/d$V$ maps reported in Ref.~\cite{Cochrane_2015}.

\subsection{Detailed alignment plots for TCNE}\label{sec:SI-tcne}

\begin{figure*}
    \centering
    \includegraphics[width=0.85\linewidth]{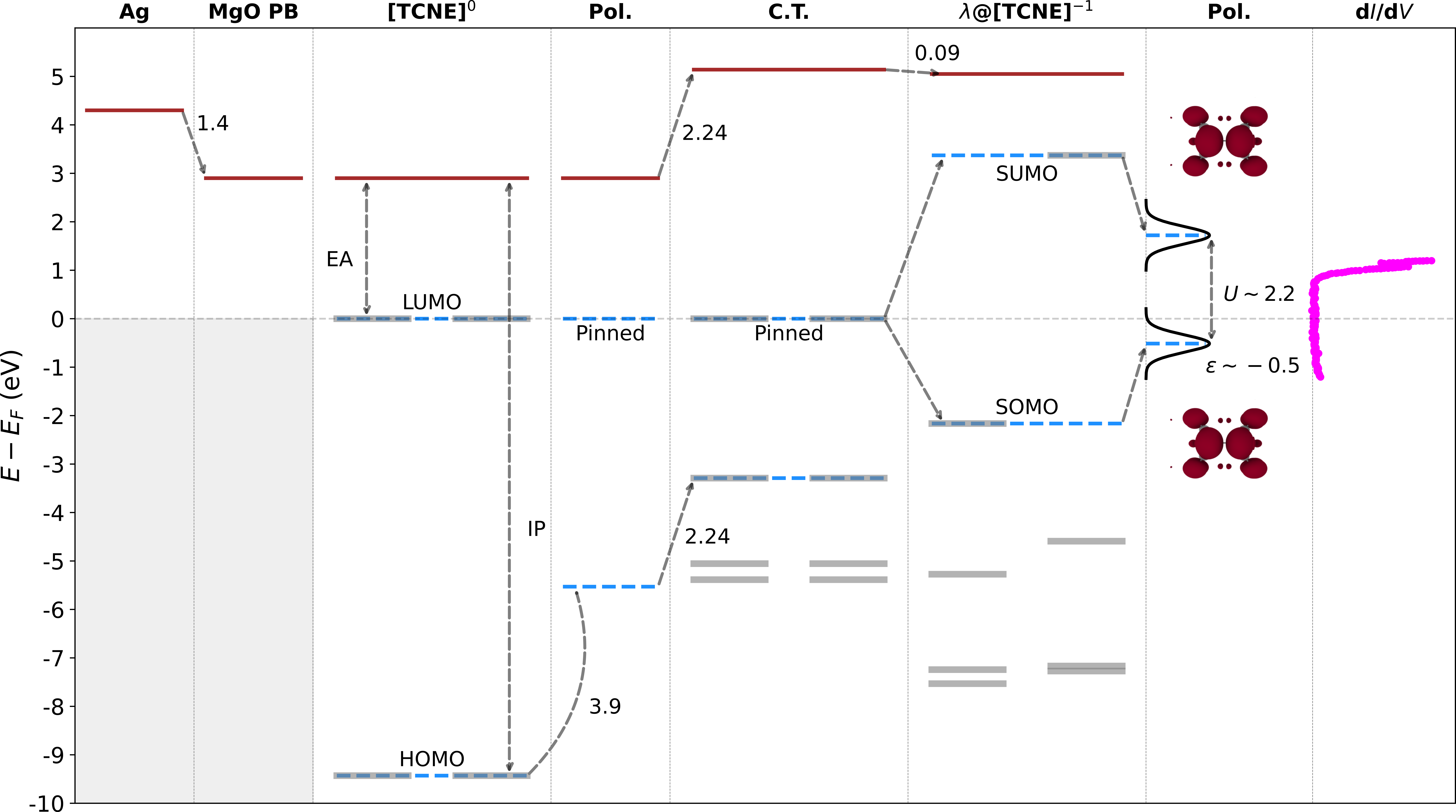}
    \caption{\textbf{Energy level alignment of TCNE on MgO/Ag.} The schematic from left to right correspond to all the effects such as Pauli pushback ($\Delta\phi_\textrm{PB}$), polarization (Pol.), charge transfer (C.T.), and reorganization of energy levels. Molecular orbital densities of the SOMO and SUMO appear as red isosurfaces. Experimental data of Ref. \cite{Wegner_2007}. }
    \label{fig:tcne}
\end{figure*}

TCNE (C$_2$(CN)$_4$) is a closed-shell molecule characterized by a large ionization potential (IP$=12.33$~eV) and a substantial electron affinity (EA$=3.51$~eV), reflecting its strong electron-acceptor character. The LUMO of TCNE immediately gets pinned to the Fermi level of the MgO/Ag substrate which indicates electron transfer from the substrate to the molecule. The small localization of the charge in the molecule causes a strong interface dipole formation. This causes a stronger upward shift of $\Delta\phi = 2.24$~eV of the vacuum level (see Fig.~\ref{fig:tcne}) compared to previous systems, such as pentacene/MgO/Ag. The [TCNE]$^{-1}$ molecule has an unpaired highly localized spin~\cite{Wegner_2007, Zhang_2020}. After the charge transfer, the [TCNE]$^{-1}$ has an IP$=7.21$~eV and EA$=1.68$~eV. By accounting for the inner relaxation energy ($\lambda$) and the polarizability, we obtain SUMO and SOMO orbitals at $E'_\mathrm{SOMO}= -0.5$ and $E'_\mathrm{SUMO}= 1.7$~eV, respectively. These values are consistent with the energy positions inferred from STS maps reported in Ref.~\cite{Wegner_2007}.

\subsection{Detailed alignment plots for Ti}\label{sec:SI-ti}

Gas-phase Ti exhibits an ionization potential and electron affinity of 7.89 eV and -0.86 eV, respectively. When aligned with Ag/MgO, neither level lies close to the Fermi energy. Even after including substrate polarization ($2P = 6.73$~eV), neither of the frontier orbitals ORB$_1$ and ORB$_2$ approaches the Fermi level. This suggests the absence of charge transfer and a $\Delta\phi=0.71$~eV in the Schottky Mott limit. 

In contrast, explicit calculations for the full Ti/MgO/Ag system reveal charge transfer from the adsorbate to the substrate, corresponding to [Ti]$^{+1}$, consistent with recent spin-excitation measurements of Ti on MgO/Ag \cite{Phark2026}. The short Ti–O distance (1.8 \AA) further suggests significant hybridization between Ti orbitals and the oxide layer, consistent with a chemisorbed configuration. Altogether, these findings indicate a breakdown of the energy-level alignment rules, emphasizing that single-atom adsorbates do not follow the same behavior predicted by our model for physisorbed systems.

\bibliography{Reference}

\end{document}